\documentclass[twocol]{arxiv}
\pdfoutput=1

\journal{jamc}
\usepackage{xfrac}
\usepackage{booktabs}
\usepackage{verbatim}
\usepackage{bm}
\usepackage{anyfontsize}
\usepackage[normalem]{ulem}
\usepackage{float}
\usepackage{amsmath}
\usepackage{amssymb}
\usepackage{natbib}
\usepackage{hyperref}

\newcommand{\ra}[1]{\renewcommand{\arraystretch}{#1}}

\bibpunct{(}{)}{;}{a}{}{,}

\def\x{{\mathbf{x}}}

\title{Unlocking GOES: A Statistical Framework for Quantifying the Evolution of Convective Structure in Tropical Cyclones}

    \authors{Trey McNeely\correspondingauthor{Trey McNeely, 
     Carnegie Mellon University Department of Statistics \& Data Science, 
     4909 Frew St, Pittsburgh, PA 15213.} and Ann B. Lee}

     \affiliation{Department of Statistics \& Data Science, Carnegie Mellon University, 
     Pittsburgh, Pennsylvania}

\email{imcneely@andrew.cmu.edu}

    \extraauthor{Kimberly M. Wood}
    \extraaffil{Department of Geosciences, Mississippi State University, Mississippi State, Mississippi}

    \extraauthor{Dorit Hammerling}
    \extraaffil{Department of Applied Mathematics and Statistics, Colorado School of Mines, Golden, Colorado}

\abstract{\fontsize{12}{11}\selectfont Tropical cyclones (TCs) rank among the most costly natural disasters in the United States, and accurate forecasts of track and intensity are critical for emergency response. Intensity guidance has improved steadily but slowly, as processes which drive intensity change are not fully understood. Because most TCs develop far from land-based observing networks, geostationary satellite imagery is critical to monitor these storms.  However, these complex data can be challenging to analyze in real time, and off-the-shelf machine learning algorithms have limited applicability on this front  due to their ``black box'' structure. This study presents analytic tools that quantify convective structure patterns in infrared satellite imagery for over-ocean TCs, yielding lower-dimensional but rich  representations that support analysis and visualization of how these patterns evolve during rapid intensity change. The proposed \texttt{ORB} feature suite targets the global \texttt{O}rganization, \texttt{R}adial structure, and \texttt{B}ulk morphology of TCs. By combining \texttt{ORB} and empirical orthogonal functions, we arrive at an interpretable and rich representation of convective structure patterns that serve as inputs to machine learning methods. This study uses the logistic lasso, a penalized generalized linear model, to relate predictors to rapid intensity change. Using \texttt{ORB} alone, binary classifiers identifying the presence (versus absence) of such intensity change events can achieve accuracy comparable to classifiers using environmental predictors alone, with a combined predictor set improving classification accuracy in some settings. More complex nonlinear machine learning methods did not perform better than the linear logistic lasso model for current data.
}

\hypersetup{draft}
\begin{document}

\maketitle

\section{Introduction}\label{sec:introduction}

Tropical cyclones (TCs) can be long-lived, powerful, and spatially extensive; these three factors place them among the most damaging and costly natural disasters in the United States \citep{Klotzbach2018}. Our TC guidance has improved over the past few decades, though reductions in track forecast error have outpaced reductions in intensity forecast error \citep{DeMaria2014}. Rapid intensity change ($\geq$30-kt increase or decrease in 24 h) continues to challenge forecasters, and the need to better understand the processes which drive these changes spurred the Hurricane Forecast Improvement Project (HFIP\footnote{\url{http://www.hfip.org/}}) with specific goals to further reduce both track and intensity forecast errors.

Because the near-storm environment influences TC intensity, understanding and predicting the evolution of that environment is a key component in forecasting intensity change. The operational Statistical Hurricane Intensity Prediction Scheme (SHIPS; \citealt{DeMaria1999}) uses large-scale, numerical weather prediction (NWP)-derived diagnostics of environmental factors known to affect TC intensity change, including vertical wind shear and relative humidity. SHIPS also relies on observations such as infrared (IR) imagery and sea surface temperature (SST). Environments in which TCs preferentially undergo rapid intensification (RI; operationally defined as an increase of $\geq$30 kt in 24 h) tend to have higher SSTs, greater atmospheric humidity, and lower vertical wind shear (e.g., \citealt{kaplan2003large}). To target RI events, a corresponding subset of SHIPS predictors was used to develop the SHIPS Rapid Intensification Index (SHIPS-RII; \citealt{Kaplan2010}). SHIPS predictors can also provide insight into the characteristics of rapid weakening events (RW; defined as a decrease of $\geq$30 kt in 24 h; \citealt{wood2015definition}).

Though NWP-derived predictors are included in SHIPS, global NWP models such as the Global Forecast System (GFS) cannot resolve subgrid-scale processes like convection. \emph{In situ} TC observations aid in capturing small-scale storm features, but such aircraft reconnaissance is infrequent and expensive. Because most TCs develop and strengthen far from dense, land-based observing networks, analysts have long relied on geostationary satellite observations to monitor TCs. Since colder cloud tops imply deeper convection and thus stronger updrafts, IR observations of cloud-top brightness temperatures ($T_b$) provide a proxy for convective strength. In addition, as TCs intensify, the distribution of convection becomes more symmetric around the storm center (i.e., axisymmetric). The relationship between convective organization and the maximum intensity of TCs was used to develop the original Dvorak technique \citep{Dvorak1975}, a subjective method relating patterns in IR imagery to TC intensity.

Today, the Advanced Baseline Imager (ABI; \citealt{schmit2017closer}) on Geostationary Operational Environmental Satellite (GOES)-16 and -17 observes the North Atlantic (NAL) and eastern North Pacific (ENP) every 10 minutes in 16 bands at spatial resolutions from 2 km for infrared channels to 0.5 km for the visible Band 2. The current version of SHIPS incorporates ABI IR observations as percent coverage of $T_b$ below -30°C and standard deviation of $T_b$, each computed within a 50-200-km annulus centered on the TC \citep{kaplan2015rii}. However, single-value representations produced every 6 h cannot fully utilize the information provided by these finer spatial and temporal resolutions. How, then, do we objectively ``unlock'' the increasingly rich information contained in geostationary IR imagery that may highlight physical processes associated with short-term ($\leq$24 h) TC intensity change? From a practical point of view, there are two challenges: i) forecasters have access to increasingly detailed data but have the same limited time (6 hours) to prepare each operational forecast, and ii) although machine learning algorithms theoretically can analyze sequences of high-resolution images in an automated way, the extracted features can be difficult to interpret. However, if we can extract meaningful features that are both informative inputs to machine learning methods and interpretable by forecasters, then we can reduce forecaster effort and leverage some of the predictive power of modern statistical methods.

To address these challenges, we develop a rich and scientifically motivated family of IR imagery descriptors through our \texttt{ORB} feature suite for $T_b$. These features identify patterns in $T_b$ structure in three broad categories: (i) global \texttt{O}rganization, (ii) \texttt{R}adial structure, and (iii) \texttt{B}ulk morphology.  Each category targets one aspect of convective structure by mapping the complex $T_b$ image data to summary statistics computed over a range of temperature or radial distance thresholds. We will hereafter refer to the continuous treatment of \texttt{ORB} statistics as \texttt{ORB} functions. As demonstrated in Figures~\ref{fig:deviationangles} and~\ref{fig:skewsize}, these functions transition from threshold-based statistics (e.g., coverage of $T_b$ below -30$^\circ$C) to continuous functions of thresholds (e.g., coverage of $T_b$ below a continuously varying temperature). Furthermore, we develop an empirical orthogonal function (EOF) representation of each \texttt{ORB} function to capture its main modes of variability. The EOFs are computed separately for the NAL and ENP basins via principal component analysis (PCA); see Appendix~\ref{app:pca} for details. 

Our final representation of TC convective structure is physically interpretable (Figures~\ref{fig:radialbasis}-\ref{fig:sizebasis}) and supports both analysis and visualization of evolving convective structure patterns in near-real time (Section~\ref{sec:methods}\ref{sec:functional}). The \texttt{ORB} functions capture some IR patterns analogous to the advanced Dvorak technique (ADT; \citealt{olander2007advanced,olander2019advanced}) scene types as well as novel approaches to potentially informative convective patterns. For example, the EOFs for radial profiles and bulk morphology can mimic the ADT’s cloud region and eye region scene type scores, whereas global organization provides an alternate approach to assessing $T_b$ axisymmetry versus the cloud symmetry value included in the ADT. The \texttt{ORB} framework is also extensible to any thresholded feature, as defined in Section~\ref{sec:methods}; this enables further characterization of both additional ADT scene types and novel patterns in an objective, interpretable fashion.

This study focuses on applying the \texttt{ORB} framework to develop interpretable and rich descriptions of $T_b$ structure. To demonstrate that this suite of features retains information relevant to short-term intensity change, we build statistical models to diagnose the current state of TC intensification. We construct separate binary classifiers for RI and RW events that ask ``Is this TC currently in the midst of an RI event (or RW event)?” To predict the probability that the current state meets RI or RW criteria, we use logistic lasso, a penalized logistic regression that automatically selects a set of relevant predictors by shrinking the regression coefficients of the other (i.e., irrelevant) predictors to zero. For the predictors included in this study, we find that several harder-to-interpret, nonlinear machine learning methods result in similar performance as lasso, and thus we focus on the latter given its interpretability.

Our analysis indicates that a combination of \texttt{ORB} features and the lasso can provide a powerful observational tool for relating short-term intensity change to convective structure patterns in over-ocean TCs. These features i) enable rapidly digested summary and visualization of evolving convective structures while ii) capturing the complexity of those structures across many thresholds in a rich and extensible framework. We find that binary classification methods restricted to GOES-derived \texttt{ORB} predictors  achieve an accuracy at least comparable to classification methods restricted to SHIPS environmental predictors when distinguishing rapid change from non-rapid change events. Adding the \texttt{ORB} suite to the environmental predictor set can improve the results beyond individual predictor sets, with RI/non-RI in the NAL basin showing significant promise (Figure~\ref{fig:AUC_comparison} and Table~\ref{tab:pvals}). Finally, \texttt{ORB} allows us to interpret regression coefficients in a way that enables physical insight into the models (Figure~\ref{fig:lasso_coef_NAL}).

\section{Data}\label{sec:data}
\begin{figure}
    \centering
    \includegraphics[width=\linewidth]{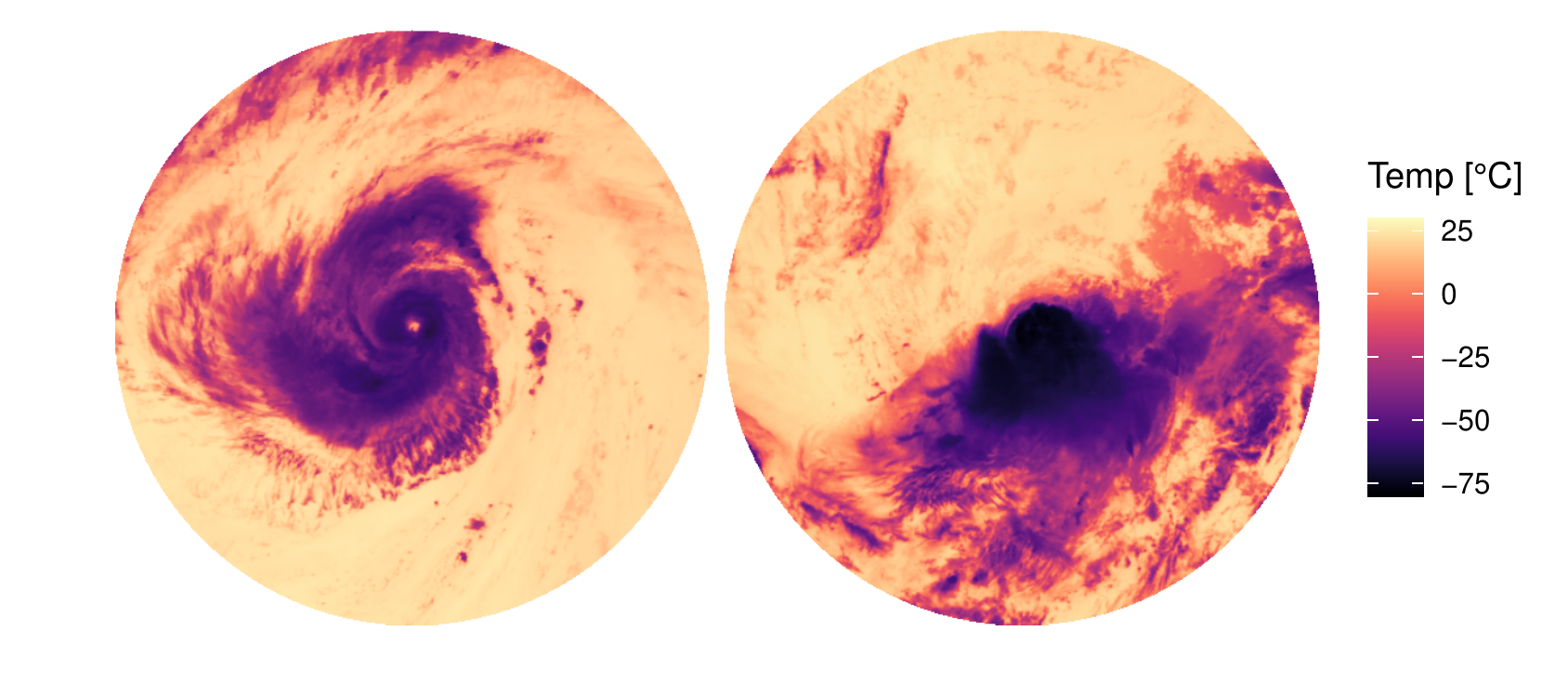}
    \caption{GOES $T_b$ stamps for Edouard at 18 UTC 16 Sept 2014 (left; showing an eye-eyewall structure at an intensity of 95 kt) and for Nicole at 1 UTC 9 Oct 2016 (right; showing an asymmetric central dense overcast at an intensity of $\sim$47 kt).
    }
    \label{fig:nicole_edouard}
\end{figure}

The GOES-East/West satellites provide $T_b$ imagery with high and consistent spatial and temporal resolutions. NOAA's GridSat-GOES database contains historical, hourly GOES Channel 4 (IR band) $T_b$ remapped to an even 0.04$^\circ$ grid over the full GOES domain (75$^\circ$N to 75$^\circ$S and 150$^\circ$E to 5$^\circ$W) in the Western Hemisphere through GOES-15 (up to 2017 at the time of this study; \citealt{Knapp2018}). Though these resolutions are coarser than the ABI, GridSat-GOES covers the period 1994-2017 and thus ensures we have a reasonably large data set for analysis. Due to early data acquisition difficulties, this study uses the period 1998-2016; future work will use the full period 1994-present. We refer to the TC-centered field at a given time as a ``stamp.'' This term is borrowed from observational astronomy, where small digital images of the sky centered around an astronomical object are sometimes referred to as postage stamps. The earliest versions of \texttt{ORB} were inspired by nonparametric approaches to quantifying galaxy morphology (e.g., \citealt{conselice2014evolution}).

We use the National Hurricane Center's Hurricane Database (HURDAT2; \citealt{Landsea2013}) to limit our TC sample to tropical time steps with intensities of at least 50 kt to ensure identifiable structure in $T_b$ and exclude extratropical times. From these 6-h samples, we compute 24-h intensity change to identify rapid vs non-rapid change events and then interpolate each track to hourly resolution to extract the associated GridSat-GOES $T_b$ stamp of radius 800 km (Figure~\ref{fig:nicole_edouard}).
\begin{table}[t]
\ra{1.2}
\centering
\begin{tabular}{@{}l|rr|r@{}}\toprule
                                  & NAL            & ENP    & Total                  \\ \hline
6-h HURDAT2 entries                & 8,438          & 8,080  & 16,518                 \\ 
$\geqslant 50$-kt HURDAT2 entries  & 4,111          & 3,400  & 7,511                  \\ 
24-h history available            & 4,017          & 3,339  & 7,356                  \\ 
$\geqslant 250$ km from land       & 2,225          & 2,627  & 4,852                  \\ 
\textbf{GOES + SHIPS available}   & \textbf{1,236} & \textbf{1,575} & \textbf{2,811} \\
RI Observations                    & 361            & 602    & 963                    \\ 
RW Observations                    & 221            & 587    & 808                    \\ \hline
Unique TCs                        & 154            & 206    & 360                    \\ \hline
RI Events                          & 71             & 103    &  174                     \\
RW Events                          & 56             & 106    & 162                     \\
\bottomrule
\end{tabular}
\caption{\emph{Observations used in our study:} The top rows of the table refer to numbers of stamps. The middle row indicates the number of TCs in the sample. The bottom rows indicate the number of RI/RW events, which each consist of multiple temporally-adjacent stamps. The statistical models in Section~\ref{sec:models} are based on the final \textbf{GOES + SHIPS available} sample. The sample size in the NAL basin is reduced due to more frequent landfall than in the ENP basin.}
\label{table:data}
\end{table}

\begin{table*}[t]
\ra{1.2}
\centering
\begin{tabular}{@{}p{.6cm}p{2.75cm}p{.005cm}p{.85cm}p{4.35cm}p{.005cm}p{.9cm}p{2.5cm}@{}}\toprule
\multicolumn{8}{c}{Predictor Sets and Variable Descriptions}\\
\midrule
\multicolumn{5}{c}{SHIPS~+~Persistence ($d=3+48=51$)} &&& \\
\cmidrule{1-5}
&&& \multicolumn{5}{c}{SHIPS~+~\texttt{ORB} ($d=48+68=116$)} \\
\cmidrule{4-8}
\multicolumn{2}{c}{Persistence ($d=3$)} && \multicolumn{2}{c}{SHIPS-only ($d=12\times4=48$)} && \multicolumn{2}{c}{\texttt{ORB}-only ($d=17\times4=68$)}\\
\cmidrule{1-2} \cmidrule{4-5} \cmidrule{7-8}
$V$ & Current intensity && SHRD & 850--200-hPa shear magnitude {(200-800 km)} && DAV & 3 \texttt{ORB} coefficients, Section~\ref{sec:methods}\ref{sec:orb}(\ref{sec:organization})\\
$\Delta_{6}V$ & 6-h intensity change && SHDC & 850--200-hPa vortex-removed shear magnitude, (0-500 km) && RAD & 3 \texttt{ORB} coefficients, Section~\ref{sec:methods}\ref{sec:orb}(\ref{sec:radial})\\
$\Delta_{12}V$ & 12-hour to 6-h intensity change && SHRS & 850--500-hPa shear magnitude && SIZE & 2 \texttt{ORB} coefficients, Section~\ref{sec:methods}\ref{sec:orb}(\ref{sec:bulk})\\
&&& OHC & Ocean heat content from altimetry where available, from SST anomaly otherwise  && SKEW & 3 \texttt{ORB} coefficients, Section~\ref{sec:methods}\ref{sec:orb}(\ref{sec:bulk})\\
&&& RSST & Reynolds SST && SHAPE & 3 \texttt{ORB} coefficients, Section~\ref{sec:methods}\ref{sec:orb}(\ref{sec:bulk})\\
&&& RHLO & 850--700-hPa relative humidity, (200-800 km) && ECC & 3 \texttt{ORB} coefficients, Section~\ref{sec:methods}\ref{sec:orb}(\ref{sec:bulk})\\
&&& RHMD & 700--500-hPa relative humidity, (200-800 km) && $\Delta_{6}(\cdot)$ & 6-h change in the above predictors\\
&&& RHHI & 500--300-hPa relative humidity, (200-800 km) &&$\Delta_{12}(\cdot)$ & 12-hour change in the above predictors\\
&&& VMPI & Maximum potential intensity from Kerry Emanuel equation &&$\Delta_{24}(\cdot)$ & 24-h change in the above predictors\\
&&& U200 & 200-hPa zonal wind, (200-800 km) &&&\\
&&& LAT & Latitude, $^\circ$N from equator &&&\\
&&& LON & Longitude, $^\circ$E from prime meridian &&&\\
&&& $\Delta_{6}(\cdot)$ & 6-h change in the above predictors &&&\\
&&& $\Delta_{12}(\cdot)$ & 12-hour change in the above predictors &&&\\
&&& $\Delta_{24}(\cdot)$ & 24-h change in the above predictors &&&\\
\bottomrule
\end{tabular}
\caption{Predictors in each predictor set. Lags for persistence ($\Delta_6V$ and $\Delta_{12}V$) are consistent with SHIPS persistence predictor INCV, the incremental 6 hour changes in intensity. Note that the use of three different lags will multiply the total number of predictors for each of SHIPS-only and \texttt{ORB}-only by a factor of four.}
\label{tab:predictors}
\end{table*}

The SHIPS developmental database provides historical values of observed and NWP-derived predictors used to generate SHIPS forecasts. In this study, we use SHIPS predictors for reasons of availability and known correlation with TC intensity change, not for the purpose of comparing our statistical model with SHIPS operational models. Our goal is to classify the current (rather than the future) state of the TC as RI/non-RI (or equivalently RW/non-RW). We restrict our analysis to the 0-h SHIPS values; that is, we only use values valid at each time step rather than the forecast values valid at subsequent times. We are then able to demonstrate the merits of the \texttt{ORB} feature suite in interpretable analyses when used alongside selected 0-h SHIPS predictors. We include two observed oceanic SHIPS predictors: ocean heat content (OHC) and Reynolds sea surface temperature (RSST). We also select eight NWP-derived atmospheric fields: 200-hPa zonal wind (U200), relative humidity at three different levels (RHLO, RHMD, and RHHI), three measures of vertical wind shear (SHRD, SHDC, and SHRS), and maximum potential intensity (VMPI) \citep{SHIPSdev}. The predictors in this initial set are selected for their i) relationship to TC convection, ii) suspected relevance to current intensity given our study's goal of classifying current intensity change, and iii) interpretability. For example, 200--850-hPa vertical wind shear calculated over a 200-800-km annulus (SHRD) and over a 0-500-km annulus (SHDC) are part of the SHIPS-RII predictor suite \citep{kaplan2015rii}, and 500--850-hPa vertical wind shear (SHRS) also correlates with 24-h intensity change \citep{rhome2006calculation}. These eight NWP-derived atmospheric predictors and two oceanic predictors are used alongside our $T_b$-derived predictors to assess the value of combining environmental predictors with TC structure information (Table \ref{tab:predictors}) in predicting the current rate of TC intensity change.

In addition to a minimum intensity of 50 kt, we further restrict our sample to over-water TCs to reduce the influence of land on TC convective structure. For a TC center to be considered over water, it must i) be at least 250 km from land and ii) not move within 250 km of land during the 24-h window under consideration. Since rapid intensity changes of at least 30 kt in 24 h are rare events, we ensure our study relies on a reasonable sample size by defining rapid change events as a 24-h intensity change of at least 25 kt; this lower threshold is frequently examined in operational forecasts such as SHIPS-RII as well as RI/RW literature \citep{Kaplan2010,kaplan2015rii,wood2015definition}. We examine the evolution of all TCs that meet our criteria between 1998 and 2016 in the NAL and ENP basins, and we analyze both intensification and weakening events. Finally, we only include stamps with less than 5\% of pixels missing.

The above criteria applied to 1998-2016 produce a data set of 2,811 6-h observations (Table~\ref{table:data}), including 174 distinct RI events and 162 distinct RW events---that is, non-overlapping runs of consecutive 6-h time steps during which a TC is undergoing RI or RW. While the above sample is used to analyze RI and RW events, we will compute the features described in Section~\ref{sec:methods} on nearly 25,000 hourly stamps (14,470 NAL, 9,818 ENP) to track the evolution of TCs at finer temporal resolution; that is, we develop our \texttt{ORB} feature suite using all stamps from the time the TC first breaches the 50-kt threshold until the last time it drops below 50 kt, regardless of proximity to land. Due to the relaxed restrictions on the included stamps, this set of hourly stamps is more than six times larger than the restricted, 6-hr set used for the modeling of Section~\ref{sec:models}.

\section{\texttt{ORB} Features for Convective Structure}\label{sec:methods}

\texttt{ORB} seeks to support statistical analysis that includes interpretable $T_b$ features derived from IR imagery. This section describes the development of ``\texttt{ORB} coefficients'' (which later serve as inputs to our statistical models) from a suite of \texttt{ORB} statistics. Throughout this section, we denote the brightness temperature at location $\mathbf{s}$ as $T_b(\mathbf{s})$. Each \texttt{ORB} statistic is based on a single $T_b$ stamp and is parameterized by a threshold value for either the maximum cloud-top temperature considered, $c$, or the radial distance from the TC center, $r$. Though all of these statistics are functions of both $T_b$ and either $c$ or $r$, we will for notational convenience simply denote them by $f(c)$ and $f(r)$. The values of $c$ and $r$ can be fixed, or they can be allowed to vary continuously, resulting in ``fixed-threshold'' \texttt{ORB} statistics or ``continuous'' \texttt{ORB} functions, respectively.

\subsection{From \texttt{ORB} Statistics to \texttt{ORB} Functions}\label{sec:orb}
\texttt{ORB} statistics quantify the spatial structure of a $T_b$ stamp by capturing i) global \texttt{O}rganization by using departures of the image gradient from perfect symmetry about the center of the storm, ii) \texttt{R}adial structure by using azimuthal averages of $T_b$ about the center, and iii) the \texttt{B}ulk morphology of the storm by using descriptions of $T_b$ level sets. These \texttt{ORB} statistics strike a balance in the trade-off between interpretability and descriptiveness. More specifically, they attempt to capture $T_b$ structure using a handful of numbers which correspond to intuitive aspects of TC structure (interpretability) while retaining as much of the information contained in the original $T_b$ image as possible (descriptiveness).

Rather than describing properties of a stamp for a fixed threshold (as in a single \texttt{ORB} statistic), we can consider the whole range of thresholds simultaneously. This is the basic idea behind \texttt{ORB} functions. For example, the deviation angle variance (DAV; defined in Section \ref{sec:methods}\ref{sec:orb}\ref{sec:organization}) is typically computed for a fixed radius (e.g., 250 km) or over a few different radii. However, we can choose any arbitrarily high number of sample thresholds. In the limit, we have a dense sampling that results in an approximation of a continuous function, the \emph{\texttt{ORB} function}.

\begin{figure}[t]
    \centering
    \includegraphics[width=.8\linewidth]{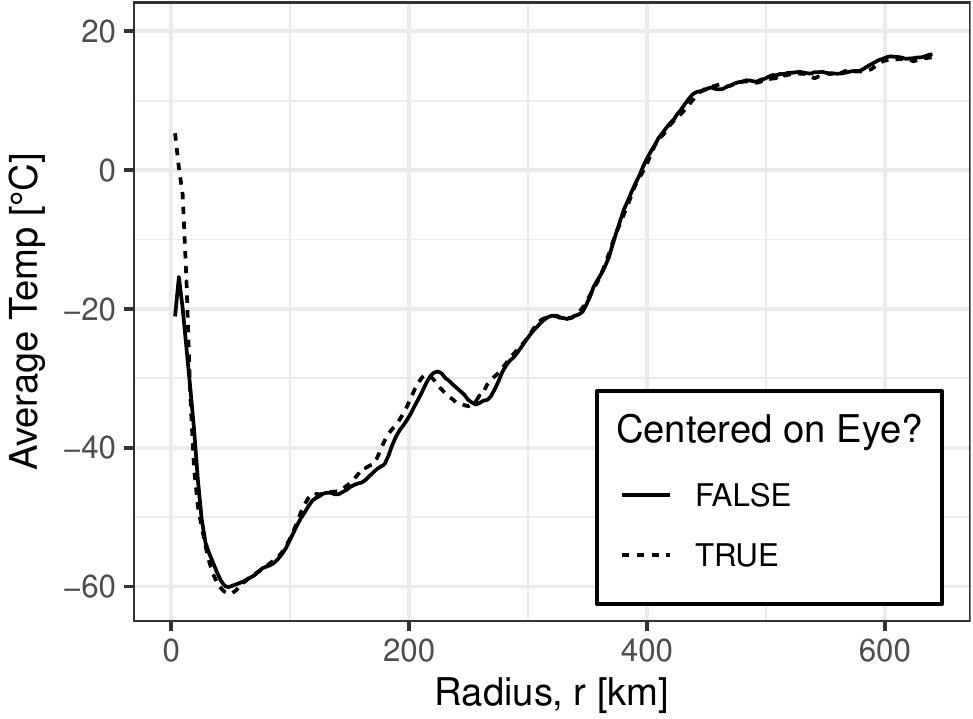}
    \caption{\emph{Radial profile $\overline{T}(r)$ as an \texttt{ORB} function for Radial structure}: Radial profiles of Edouard's $T_b$ (Figure~\ref{fig:nicole_edouard}, 18 UTC 16 Sept 2014) are shown both centered on the best track (\emph{solid}) and centered on the eye (\emph{dashed}). If we use the best-track center as the origin about which $\overline{T}(r)$ is computed, the temperature of the eye (near $r=0$) is deflated; we start including the eyewall at $r=0$ because the center of the image is not the center of the eye.}
    \label{fig:radialprofile}
\end{figure}

For Hurricane Nicole (2016), the radial profile functions 
($\overline{T}(r)$; defined in Section \ref{sec:methods}\ref{sec:orb}\ref{sec:radial}) exhibit structure that follows TC intensity (Figure~\ref{fig:radialprofile}). However, it is unclear which choice of radial threshold would best describe this structure at all time steps. Many features of these curves can vary from TC to TC or within a single TC, while others may not even appear at all time steps. For example, the radius at which the global minimum occurs is not fixed in Figure~\ref{fig:radialprofile}, while an eye (high cloud-top temperatures near $r=0$) is only occasionally present. Though we can account for many of these subtleties, generating \textit{ad hoc} features from $\overline{T}(r)$ will sacrifice the cohesiveness of a single \texttt{ORB} function. (We documented our attempt at such an approach for $\overline{T}(r)$ in \citealt{McNeely2019}.)

In this study, we consider the entire range of thresholds $c$ or distances $r$ simultaneously. This approach ensures that we do not miss structure in the feature, as selecting just a few thresholds is likely to. The density of our sampling scheme is determined by the precision of the data; here we sample at the 0.04$^\circ$ spatial resolution of GridSat-GOES for \texttt{ORB} functions $f(r)$ of radius $r$ and every 1$^\circ$C for \texttt{ORB} functions $f(c)$ of temperature threshold $c$.

\begin{figure}[t]
    \centering
    \includegraphics[width=.9\linewidth]{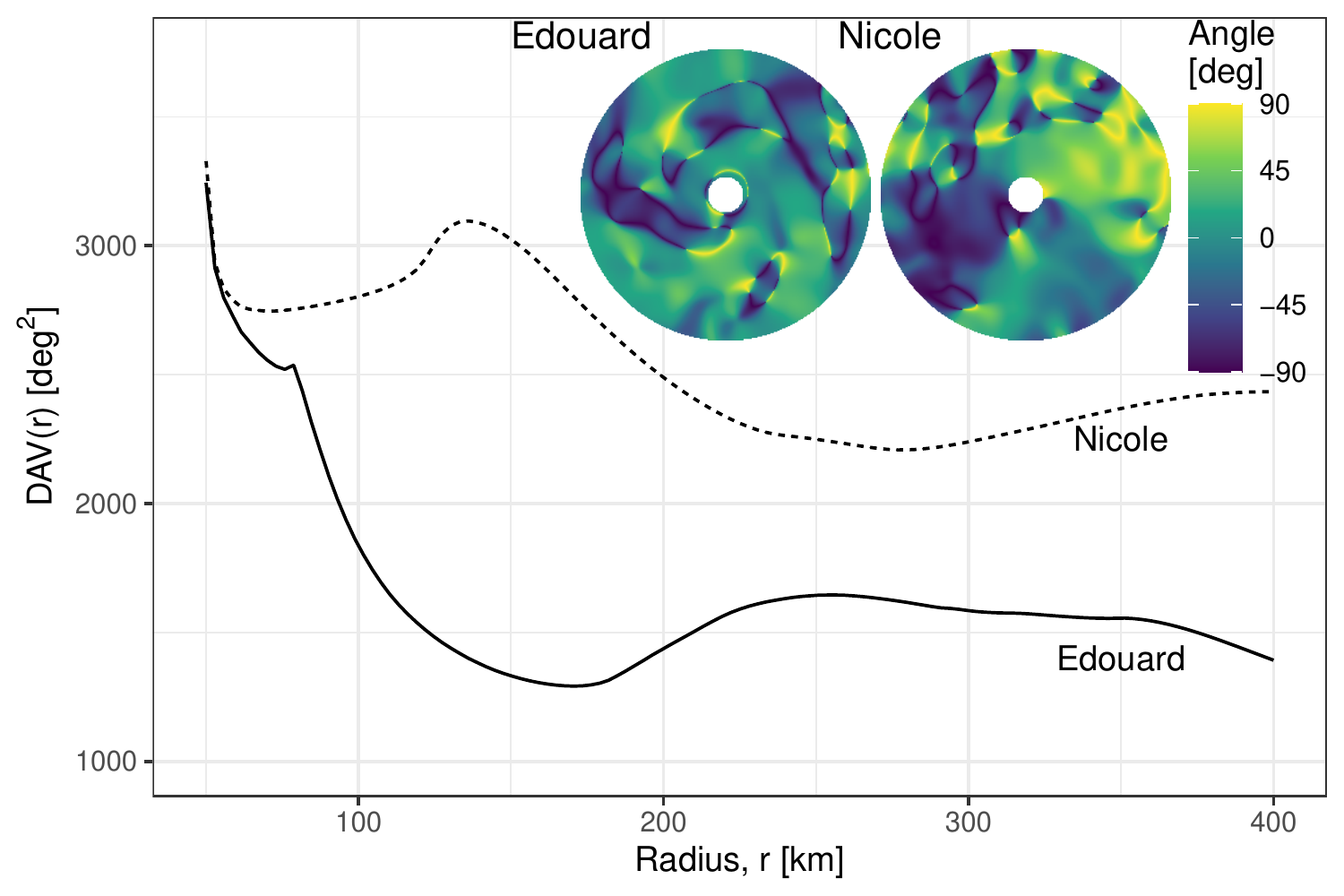}
    \caption{\emph{DAV(r) as an \texttt{ORB} function for Organization}:  The deviation angle variance (DAV) as a \texttt{ORB} function of the threshold $r$ and deviation angles for each image in Figure~\ref{fig:nicole_edouard}, Edouard (2014) (\emph{solid}, left inset) and Nicole (2016) (\emph{dashed}, right inset).}
    \label{fig:deviationangles}
\end{figure}

\subsubsection{Global \texttt{O}rganization}\label{sec:organization}

The characterization of global \texttt{O}rganization  utilizes the deviation angle variance technique \citep{Pineros2008}. We first compute the 2-D image gradient of $T_b$ at each position $\mathbf{s}=(x,y)$ (denoted $\nabla T_b(\mathbf{s})$). Consider the core structure of Edouard (2014) in Figure~\ref{fig:nicole_edouard}: the image gradient of a perfectly axisymmetric TC is expected to point either directly toward or away from the TC center. The deviation angle compares the direction of $\nabla T_b(\mathbf{s})$ to the gradient directions expected of such an idealized storm. We denote the deviation angle of the image gradient of $T_b$ at a point $\mathbf{s}$ as $\psi(\mathbf{s})$; values near $\psi(\mathbf{s})=0$ indicate local axisymmetry.

We summarize the level of organization by taking the variance of $\psi(\mathbf{s})$ over circular regions centered on the TC. We will denote the \texttt{ORB} statistic for global organization over a region of fixed radius $r$ as
\begin{align}
    \text{DAV}(r)= &\text{Var}\big[\psi(\mathbf{s})\bigm\vert\lvert\mathbf{s}\rvert\le r\big]\label{eqn:dav}
\end{align}
We will consider the value of DAV over a range of radii from $r=50$ km to $r=400$ km, resulting in \emph{\texttt{ORB} functions for global \texttt{O}rganization}. Figure~\ref{fig:deviationangles} shows DAV($r$) as \texttt{ORB} functions for the two infrared images of Figure~\ref{fig:nicole_edouard}. 

\subsubsection{\texttt{R}adial Structure}\label{sec:radial}
The characterization of \texttt{R}adial structures is based on radial profiles (e.g., \citealt{Sanabia2014}), the angular averages of $T_b(\mathbf{s})$. At a fixed distance $r$ from the TC center, define the \texttt{ORB} statistic for radial structure as
\begin{equation}
    \overline{T}(r)=\frac{1}{2 \pi}\int_{0}^{2 \pi}T_b(r,\theta)d\theta.\label{eqn:rad}
\end{equation}
The primary complication comes from centering the coordinate system (i.e., defining the location of $r=0$). \citet{Sanabia2014}  solve this by manually identifying the center of each image, whereas we \emph{automate image centering} by maximizing $\overline{T}(r)$ at low $r$ when an eye is present \citep{McNeely2019}. In this work, we use the best track center when an eye is not detected. This automated image centering is also used in the computation of DAV($r$).

Our treatment of the radial profile also differs from earlier work in its use of functional features, defined in Section \ref{sec:methods}\ref{sec:functional}, rather than designed features along the curve. \citet{Sanabia2014} identify several critical points along this curve for use as features, such as the radial location of the minimum $T_b$ within 200 km. In lieu of such designed features on $\overline{T}({r})$, Section~\ref{sec:methods}\ref{sec:functional} details a method for the recovery of data-driven functional features from general \texttt{ORB} functions; these radial profiles serve as \emph{\texttt{ORB} functions for \texttt{R}adial structure}.

Figure~\ref{fig:radial} demonstrates the relationship between the radial profile and TC intensity. Radial profiles discard angular information, making them informative primarily when bulk symmetry is high, as determined from \texttt{B}ulk morphology \texttt{ORB} functions defined in the next section. In such cases, radial profile temperatures can help discriminate between eye-eyewall structures and uniform central dense overcasts. In an eye-eyewall structure, the temperature at the center $r=0$ will be high, while the minimum of the radial profile, $\min_r \overline{T}(r)$, will be low; this typically indicates a strong TC. Conversely, a uniform central dense overcast is marked by low temperatures near the center $r=0$ due to the absence of a clear, warm eye and generally indicates a TC is below hurricane strength ($<$64 kt). The slope of the profile also provides insight into the extent of the storm's convection: gentle slopes indicate deep convection sustained over large regions, and sharp slopes indicate more localized deep convection most often associated with an eyewall.

\subsubsection{\texttt{B}ulk Morphology}\label{sec:bulk}
\begin{figure}[b]
    \centering
    \includegraphics[width=\linewidth]{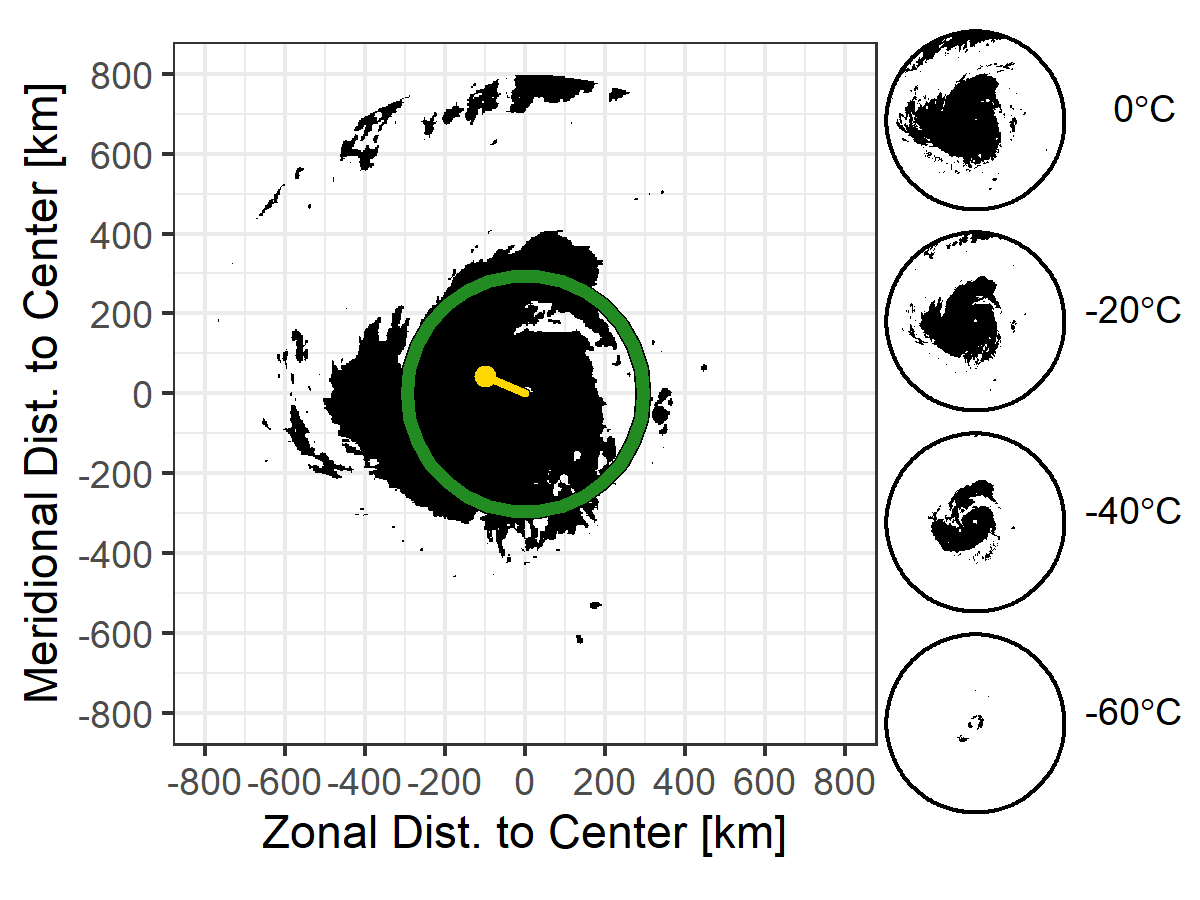}
    \caption{\emph{SIZE and SKEW as \texttt{ORB} statistics for Bulk morphology}: Example level set for a threshold of $c=-20^\circ C$ on Edouard's $T_b$ (Figure~\ref{fig:nicole_edouard}). Level sets for $\{0^\circ C, -20^\circ C, -40^\circ C, -60^\circ C\}$ are also shown (right, descending). SIZE$(c)$ is the area covered by a level set; here, the size is 400,000 km$^2$, 20\% of the stamp area. SKEW$(c)$ is the displacement of a level set's center of mass (\emph{yellow}; 34 km from center) normalized by the average displacement of points in the set (\emph{green}; 92 km from center); here, the skew is 0.37 west-northwest.}
    \label{fig:skewsize}
\end{figure}

The characterization of the \texttt{B}ulk morphology of $T_b$ utilizes sublevel sets. A sublevel set (hereafter ``level set'') on $T_b(\mathbf{s})$ identifies the points $\mathbf{s}=(x,y)$ for which $T_b$ is at or below a certain threshold. Formally, define a level set as
\begin{equation}
    \mathcal{L}(c)=\{\mathbf{s}\mid T_b(\mathbf{s})\le c\}.\label{eqn:level-set-definition}
\end{equation}
As the temperature threshold $c$ is reduced, the level set shrinks to a region of lower temperatures and typically stronger convection. This definition of $\mathcal{L}(c)$ gives rise to a variety of \texttt{ORB} statistics for bulk morphology.

\texttt{ORB} currently uses four summaries for the structure of $\mathcal{L}(c)$; see \citet{McNeely2019} for the full mathematical definitions. SIZE$(c)$, a function of the level set $\mathcal{L}(c)$, gives the area covered by $\mathcal{L}(c)$ in km$^2$ (the black region in Figure~\ref{fig:skewsize}), providing insight into the coverage of various levels of convection in a TC. Similarly, SKEW$(c)$ is a function of the level set $\mathcal{L}(c)$ and gives the displacement of the center of mass of $\mathcal{L}(c)$, normalized by the mean radius of points in $\mathcal{L}(c)$, where ``radius'' is the distance from the TC center. This reveals whether the TC cloud pattern is biased in a particular direction (Figure~\ref{fig:skewsize}). The remaining two are measures of the raggedness (SHAPE) and stretching or eccentricity (ECC) of the level set $\mathcal{L}(c)$. Note that any functions of level sets (such as our SIZE, SKEW, SHAPE, ECC) can themselves be regarded as functions of the level set threshold $c$. Hence, we refer to their continuous approximations as \emph{\texttt{ORB} functions for Bulk morphology}.

\subsection{Analysis of TC structure via \texttt{ORB} and PCA}\label{sec:functional}
\begin{figure}
    \centering
    \includegraphics[width=.9\linewidth]{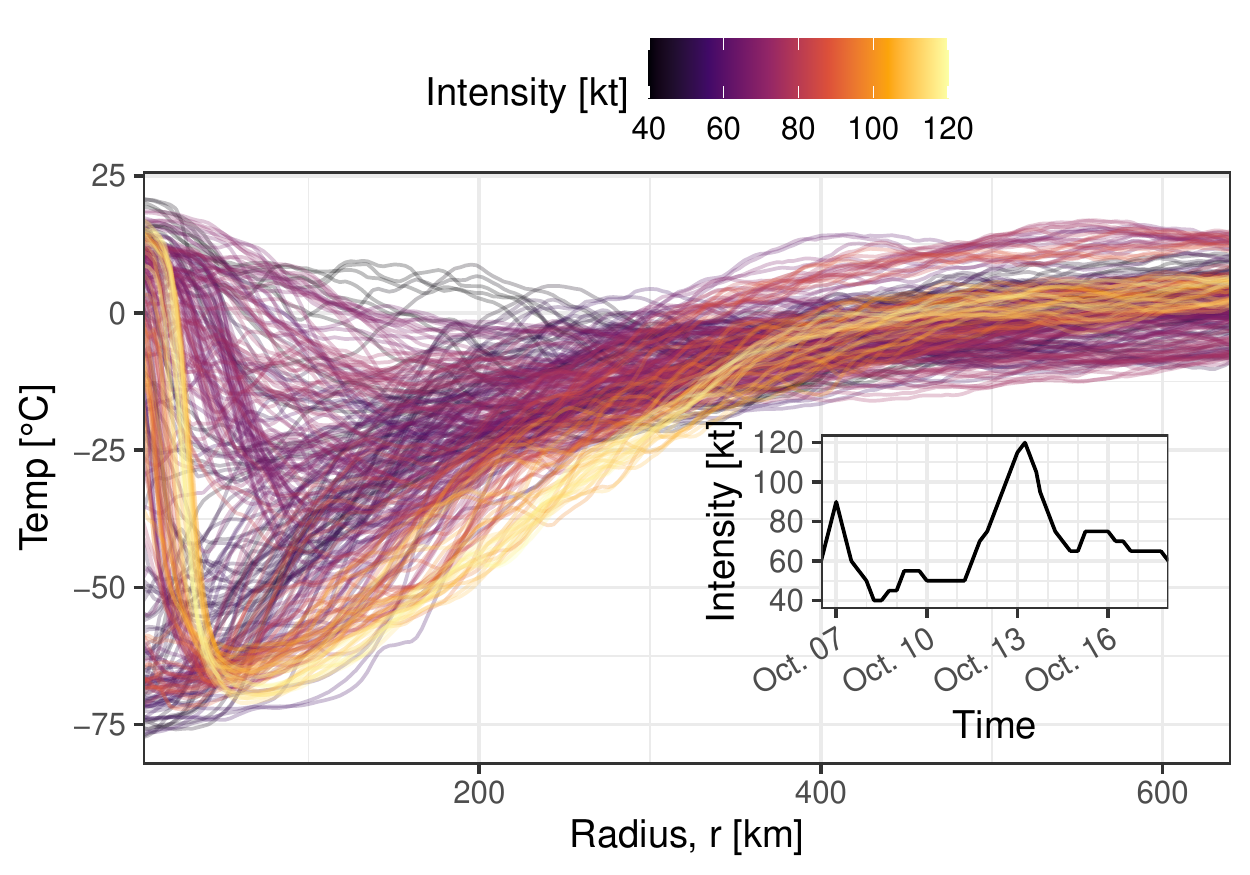}
    \caption{\emph{Hourly radial profiles:} Example hourly eye-centered radial profiles for Hurricane Nicole (2016). Lighter colors indicate higher intensities. The inset depicts Nicole's intensity trajectory. The high-intensity phase of the TC's evolution corresponds to the clearest eye-eyewall structure.}
    \label{fig:radial}
\end{figure}

{\bf Dimensionality reduction via PCA.} 
When we densely sample the \texttt{ORB} statistics, we obtain approximations of continuous \texttt{ORB} functions. Now, we will switch to a different representation (coordinate system) in which we can reduce the dimensionality of the \texttt{ORB} functions while retaining most of the information in the original functions. (Here, ``dimension'' refers to the number of values we use to represent each function; thus the original dimension of the \texttt{ORB} functions would be the number of threshold/sampling values.) A more compact representation simplifies the analysis and enables us to visualize the evolution of TC convective structure over time in a low-dimensional phase diagram defined by the new representation.

Given a suitable orthogonal basis, we can write any continuous function, such as the radial temperature profiles in Equation~\ref{eqn:rad}, as a linear combination of basis functions. More specifically, after subtracting the average over the basin, the function $f(x)$ describing deviations from the basin average is
\begin{equation}
    f(x) =  \alpha_1f_1(x)+\alpha_2f_2(x)+\cdots+\alpha_if_i(x)+\cdots,\label{eqn:basis}\end{equation}
where $f_i(x)$ denotes the i$^{th}$ basis function, and $\alpha_i=\langle f, f_i\rangle$ is the scalar orthogonal projection of $f(x)$ onto the i$^{th}$ basis function, representing how much of the shape of $f_i(x)$ is present in $f(x)$. We call the value of $f_i(x)$ the \emph{loading} of the basis function at $x$.

For continuous functions, the number of elements in a basis is infinite. However, provided that a proper basis is chosen, one can reasonably approximate functions using a finite linear combination of basis functions. In this work, we will use a data-driven approach called principal component analysis to find a low-dimensional basis representation of our \texttt{ORB} functions. PCA returns an ordered set of orthogonal and normalized vectors whose linear combinations can fully reconstruct the original vectors --- that is, the densely sampled \texttt{ORB} functions. Henceforth we will refer to the basis functions $f_1, f_2, \ldots$ from PCA as {\em empirical orthogonal functions}, to emphasize that these basis functions are empirical (data-driven) and orthogonal. We compute EOFs $\{f_i\}$ separately for each basin and each \texttt{ORB} feature. The EOFs are constant in time, whereas the projections $\alpha_i$ of the stamps onto respective EOFs $f_i$ are scalars varying with time. We refer to the latter $\alpha_i$-values as {\em \texttt{ORB} coefficients}; these coefficients serve as a time-varying representation of TC convective structure.

\begin{figure}
    \centering
    \includegraphics[width=.9\linewidth]{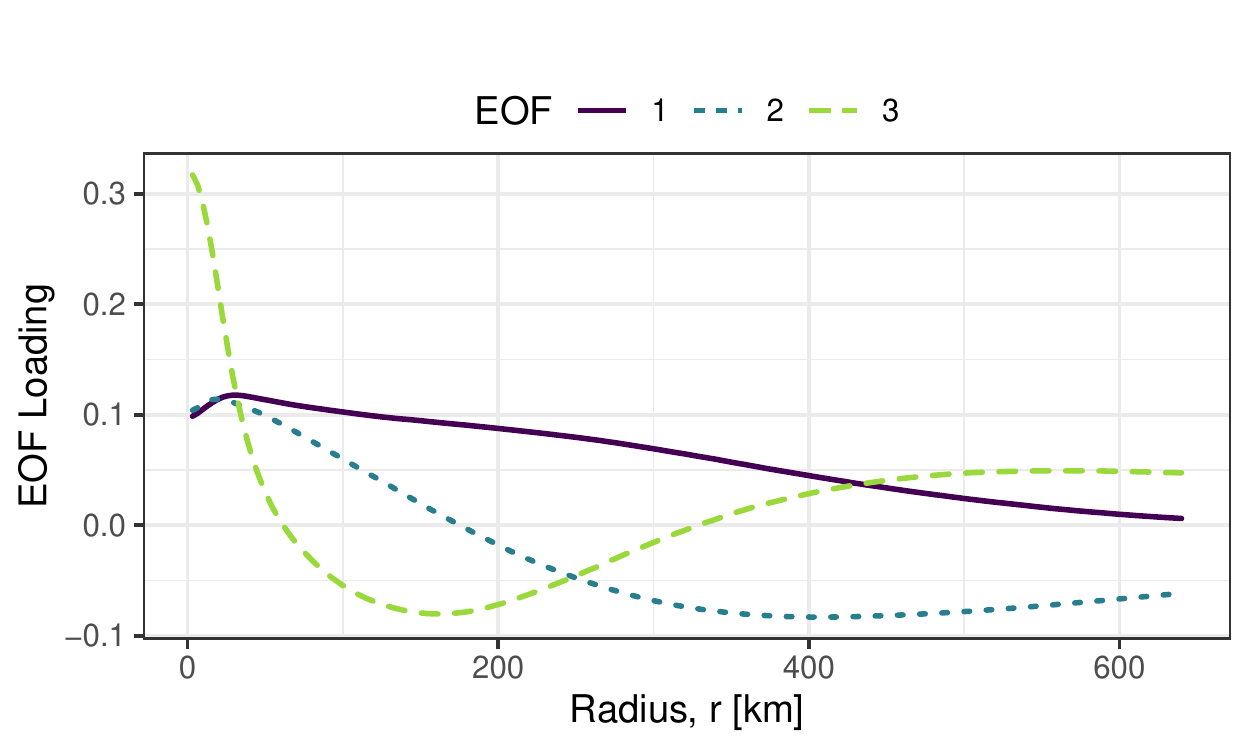}
    \caption{\emph{EOFs for radial profiles:} Results from PCA of NAL TCs. The first 3 EOFs capture 90\% of the variability in the \texttt{ORB} functions for radial structure and describe the three primary orthogonal shapes present in their profiles see Section~\ref{sec:methods}\ref{sec:functional} for details.}
    \label{fig:radialbasis}
\end{figure}

The first few EOFs provide the most information, and in our case, these tend to capture larger-scale structures of an \texttt{ORB} function. We can project an observed \texttt{ORB} function (a single function computed from a stamp) onto the first $K$ EOFs of that feature to obtain the best $K$-dimensional approximation. (Figure \ref{fig:rad_recon} demonstrates such a reconstruction for radial, RAD; Appendix~\ref{app:pca} includes details on PCA.) In our study, we use the first $K$ coefficients $\{\alpha_1,\alpha_2,...,\alpha_K\}$ of each \texttt{ORB} function at each time stamp as both inputs (predictors) to machine learning methods as well as a means to visualize the evolution of TC convective structure with time. 

{\bf Analysis of TC structure and evolution.}  We compute basin-specific EOFs separately on 14,470 NAL stamps and 9,818 ENP stamps (Section \ref{sec:data}). It only takes $K=2$ or 3 EOFs $f_i(x)$ to explain most of the variability in the six current \texttt{ORB} functions (i.e., DAV, RAD, SIZE, SKEW, SHAPE, and ECC). While objective methods (such as cross-validation) exist for selecting the number of basis functions, we here use an heuristic 90\% variance threshold. A larger set may prove useful in forecasting-only settings, but some of the higher-order basis functions lack the interpretability of the first 2-3 EOFs. 

\begin{figure}
    \centering
    \includegraphics[width=\linewidth]{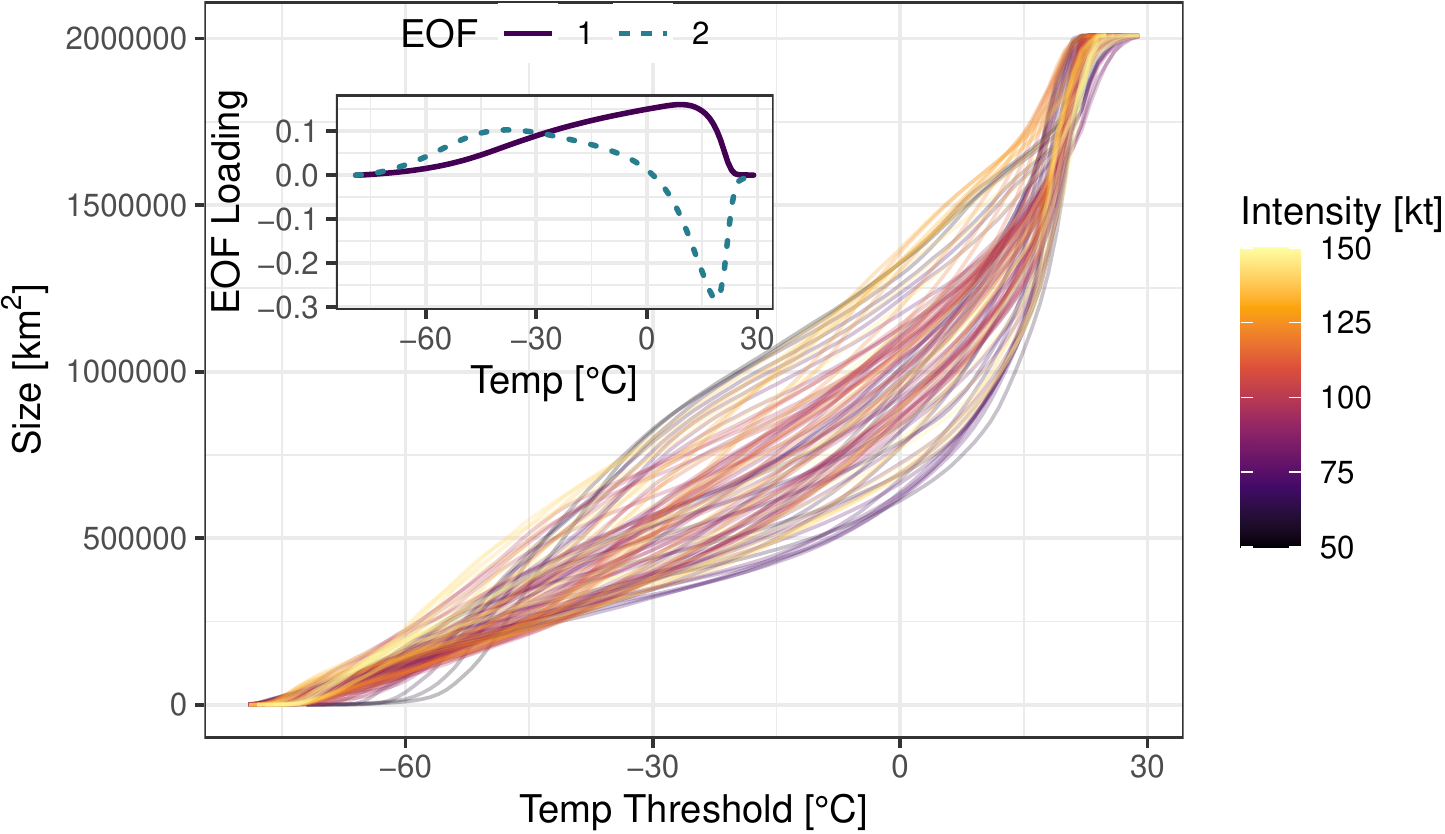}
    \caption{\emph{Hourly SIZE functions:} SIZE functions for Hurricane Katrina (2005). Lighter colors indicate higher intensities. The first two EOFs in the inset capture 94\% of the variance in the size functions.}
    \label{fig:sizebasis}
\end{figure}

Reconstruction using the first three EOFs of radial structure explains 90\% of the variance of observed radial profiles (Figure~\ref{fig:radialbasis}). EOF 1 (solid purple) roughly measures the overall temperature anomaly of the profile, with higher loadings in the TC interior; $f_1(r)$ is positive, so generally the corresponding \texttt{ORB} coefficient is positive ($\alpha_1>0$) when the radial profile $\overline{T}(r)$ has warmer-than-average cloud tops. EOF 2 (dotted blue) measures the core temperature relative to outer-band temperatures. Since $f_2(r)$ changes sign from positive to negative around 200 km, generally $\alpha_2>0$ when the radial profile shows a warmer core; hence, $\alpha_2<0$ could indicate a uniform central dense overcast or an obscured eye. Finally, EOF 3 (dashed green) indicates the presence of an eye-eyewall structure, where $f_3(r)>0$ for $r<50$ km and $r>300$ km and $f_3(r)<0$ elsewhere; hence generally $\alpha_3>0$ indicates an IR-visible eye. These three EOFs alone can construct close approximations to the wide range of the profiles apparent in Figure~\ref{fig:radial}.

In Figure~\ref{fig:sizebasis}, we apply the same \texttt{ORB} framework to the SIZE function for Hurricane Katrina (2005). For SIZE, we only need $K=2$ EOFs to capture 90\% of the variance in either basin. Because EOF 1 (solid purple) of SIZE measures the rate at which $\mathcal{L}(c)$ accumulates area with increasing temperature thresholds $c$, $\alpha_1$ for SIZE indicates colder overall $T_b$. EOF 2 (dashed green) flattens the middle of the SIZE function when $\alpha_2>0$, resulting in more coverage at high temperatures (the exposed sea surface) and low temperatures (deep convection).

\begin{figure}[t]
    \centering
    \includegraphics[width=\linewidth]{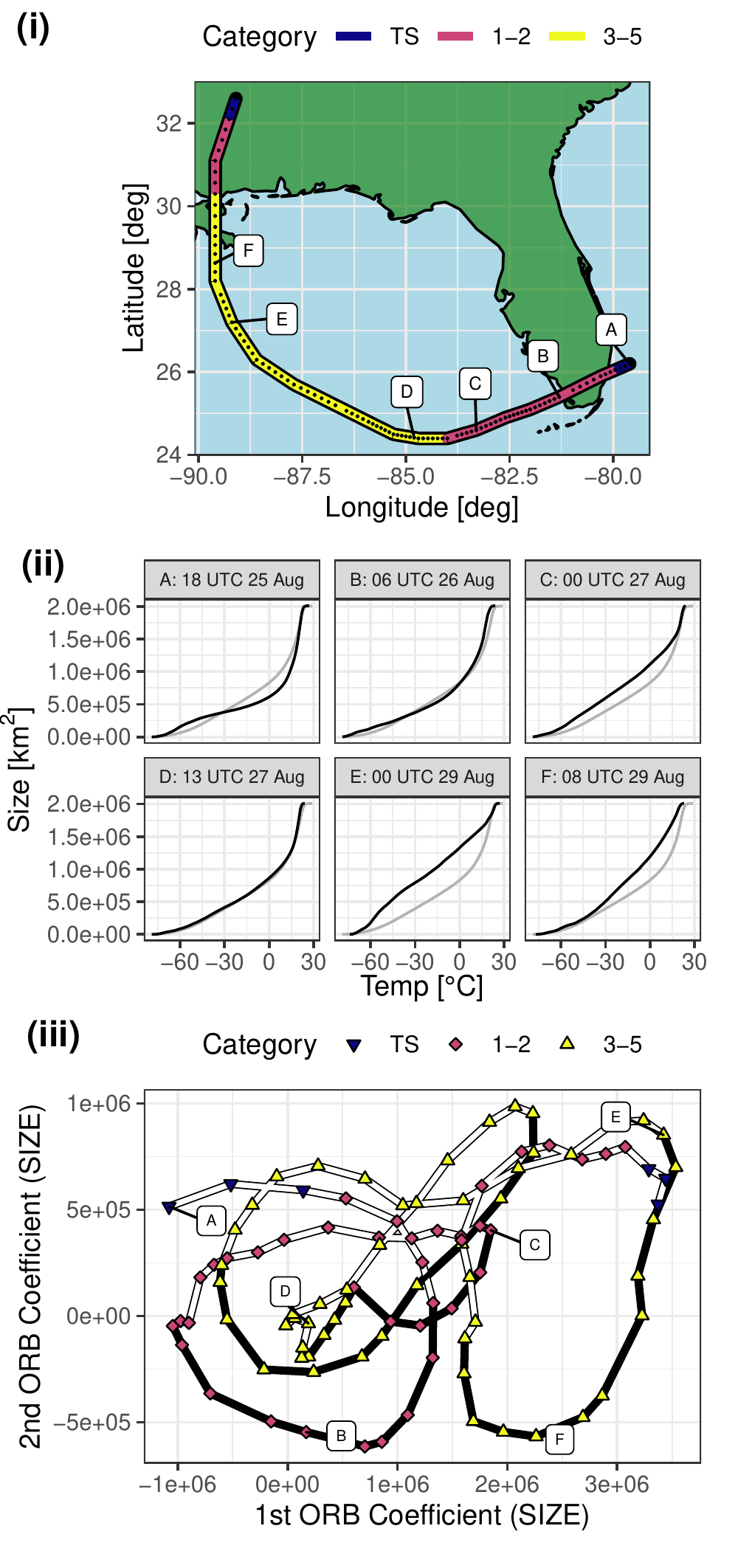}
  \caption{\emph{Visualizing SIZE evolution with \texttt{ORB}:} The spatial trajectory (\textbf{i}) shows the path of Hurricane Katrina (2005) through the Gulf of Mexico from 18 UTC 25 Aug through 00 UTC 30 Aug, with the labels marking the locations of displayed \texttt{ORB} functions. At six points (A-F, labeled) in time, we plot the SIZE function (\textbf{ii}; \emph{black}) and the sample mean SIZE function for the NAL basin (\textbf{ii}; \emph{gray}). The bottom panel  (\textbf{iii}) shows the trajectory of the \texttt{ORB} coefficients of the SIZE function in phase space during the storm's path. The clockwise cyclic path in (iii) matches the local night/day cycle as well as the storm's path in (i); see text for details.}
    \label{fig:sizetrajectory}
\end{figure}

Although this study's classification methods use 6-h values to match predictors from the SHIPS developmental database (Section~\ref{sec:models}), one benefit of $T_b$ observations is high temporal resolution. It is straightforward to track the structural evolution of TCs with \texttt{ORB} coefficients. The EOFs $f_i(x)$ for each \texttt{ORB} feature are constant within a basin, but the \texttt{ORB} coefficients are implicitly functions of time; that is, $\alpha_i=\alpha_i(t)$. One temporal phenomenon that can be observed in TCs is the diurnal cycle (e.g., \citealt{dunion2014tropical}), where deeper convection occurs near the TC core around local sunset and then spreads radially outward through the following afternoon. In Figure~\ref{fig:sizetrajectory}, the time evolution of the first two \texttt{ORB} coefficients for \texttt{SIZE} exhibits a 24-h periodic clockwise oscillation for Hurricane Katrina (2005). Though SIZE is insensitive to the location of convection and thus cannot measure the outward motion of such pulses, the observed oscillations in this phase space may be a manifestation of the TC diurnal cycle. In the phase diagram, we see two primary deviations from this cycle: when the TC turns north and accelerates (near point D, when it becomes a major hurricane [100 kt]) and when the TC begins to interact with land (the rightward translation towards point E).

\subsection{Summary of the \texttt{ORB} Framework}\label{sec:summary}

\texttt{ORB} quantifies convective structure in TCs as revealed by $T_b$ imagery in a way that is both descriptive enough to support statistical analyses, and interpretable enough to support scientific inquiry into and clear visualization of temporal evolution of rapidly changing TCs. The framework under which \texttt{ORB} is developed allows extension to other aspects of convective structure as well as other observational bands. The main steps in the construction of an \texttt{ORB} representation are as follows:
\begin{enumerate}
    \item Develop a summary of $T_b$ structure reliant on a single threshold, an \emph{\texttt{ORB} statistic}. For example, the average $T_b$ can be computed at a single distance from the TC center, such as the mapping $\overline{T}(r)$. (See Subsection~\ref{sec:methods}\ref{sec:orb} and \citealt{McNeely2019}.)
    \item Compute the summary for a dense set of threshold values. View the dense sampling as an approximation of a continuous function, the \emph{\texttt{ORB} function}. (See Figures~\ref{fig:deviationangles}, \ref{fig:radialprofile}, \ref{fig:skewsize} and Subsection~\ref{sec:methods}\ref{sec:orb}.)
    \item  Compute the \texttt{ORB} functions for all stamps from the same TC basin (here, NAL or ENP). Apply PCA to the resulting set of \texttt{ORB} functions;  the derived empirical orthogonal functions reflect the basin-specific variability of each \texttt{ORB} function. We refer to the projection of \texttt{ORB} functions onto the EOFs as \emph{\texttt{ORB} coefficients}. The \texttt{ORB} coefficients are the time-varying inputs to statistical models, as in Section~\ref{sec:models}. (See Figures~\ref{fig:radial},~\ref{fig:radialbasis}, and \ref{fig:sizebasis} and Subsection~\ref{sec:methods}\ref{sec:functional}.) 
    \item Interpret the EOFs. This interpretation relies on the design of the original \texttt{ORB} statistics (step 1). For example, PCA of the average temperature as a function of radius detects the presence of an eye-eyewall structure in its third EOF. (See Figures~\ref{fig:radialbasis},~\ref{fig:sizebasis}, and \ref{fig:sizetrajectory} and Subsection~\ref{sec:methods}\ref{sec:functional}.) 
\end{enumerate}

One can use the \texttt{ORB} coefficients to visualize or analyze TC convective evolution. In the next section, we use the \texttt{ORB} coefficients as inputs to machine learning methods that can sift through a large number of predictors and relate rapid change events to a subset of relevant predictors. The interpretability of the EOFs and associated \texttt{ORB} coefficients allows one to directly tie the results of such an analysis to meaningful aspects of convective structure.

\section{Modeling Rapid Intensity Change}\label{sec:models}

The ultimate goals of \texttt{ORB} are to i) advance scientific understanding of TC intensity change and ii) support future improvements in TC intensity forecasting by providing a framework for analysis of IR imagery. To evaluate the effectiveness of the \texttt{ORB} framework, we here consider the problem of diagnosing the state of rapid change in a TC (where RI and RW events are treated separately). The problem of obtaining the state $Y_t$ given data $X_{1:t}$ collected up to and including time $t$ is sometimes referred to as {\em filtering} \citep{cressie2015statistics}, to be distinguished from {\em forecasting} or predicting $Y_t$ based on data $X_{1:t-1}$. 

Following the operational convention of 24 hours, we label every observation in our sample as either RI or non-RI (or analogously as either RW or non-RW). If a given 6-hourly observation lies within any 24-h moving window that was classified as RI or RW according to our 25-kt definition, then that 6-hourly observation is labeled as RI or RW. All other 6-hourly observations are labeled as non-RI/non-RW.

We build a probabilistic binary classifier (see Section \ref{sec:models}\ref{sec:lasso} for details) that predicts the current state of the storm with respect to rapid change events,  using four different \emph{predictor sets} or combinations of inputs.

The four predictor sets in our analysis are:
\vspace{-.2cm}\paragraph{SHIPS-only.}  This predictor set contains only the ten SHIPS predictors we selected for this study, as well as TC latitude and longitude (see Section~\ref{sec:data} and Table~\ref{tab:predictors}).
\vspace{-.2cm}\paragraph{\texttt{ORB}-only.} This predictor set contains only \texttt{ORB} coefficients derived from $T_b$ imagery.
\vspace{-.2cm}\paragraph{SHIPS + \texttt{ORB}.} This predictor set adds \texttt{ORB} coefficients to the SHIPS-only predictor set but does not add explicit interactions between the two.
\vspace{-.2cm}\paragraph{SHIPS + Persistence.} This set includes SHIPS predictors and \emph{intensity persistence} (defined by SHIPS as the current intensity, the 6-h change in intensity, and the 12-to-6-h change in intensity). 

These predictor sets have been chosen to demonstrate the value of adding \texttt{ORB} coefficients to SHIPS environmental predictors. The fourth predictor set provides a baseline for other predictor sets or a best-case scenario of classification performance. Because RI and RW events are \emph{defined} using information partially contained in these persistence terms, one could argue that including such a term only makes sense in prognosis, but not when the goal (as in our case) is diagnosis and relating the current RI (or RW) state to meaningful predictors.

\subsection{Methods: Probabilistic Classification via Logistic Lasso}\label{sec:lasso}

We would like to answer the question: ``Is the storm currently undergoing or entering a rapid change event, and what convective features are associated with such an event?'' We proceed by constructing a statistical model --- the logistic lasso --- that uses the $d$ components of a predictor set at time $i$ (which we denote by a $d$-dimensional vector of predictors, $\x_i$) to classify the binary state $Y_i$ of the storm at that time. (For example, with the SHIPS-only predictor set in Table \ref{tab:predictors}, we have a total of $d=48$ predictors for 10 SHIPS variables plus latitude and longitude and their lagged changes, whereas $d=68$ for the \texttt{ORB}-only predictor set.) Following the previously described labeling scheme, we say that $Y_i=1$ if the $i$th observation of the TC occurs during a RI (or RW) event, whereas $Y_i=0$ corresponds to the observed absence of RI (or RW).

We will use TCs prior to 2010 to learn the {\em probability} of an ongoing rapid change event, $p_i \equiv \mathbb{P}(Y_i=1 | \x_i)$, and then use the learned statistical model to predict the probability of rapid change for TCs that occur during and after 2010; that is, our training set spans 1998-2009, while our test set spans 2010-2016.\footnote{While the statistical models use train-test data splitting, the EOFs are computed for the full data set. Such semi-supervised learning approaches, which use all data to learn a basis for dimension reduction but then predict on test data only, are common in statistical machine learning. See, for example, \cite{belkin2004semi} for classification on nonlinear manifolds.} These probabilities can be converted to binary (RI/non-RI or RW/non-RW) classes. Below, we describe the details of the logistic lasso regression model and how to properly fit and assess such a model.

{\bf Logistic lasso.} \emph{Logistic regression} is similar to traditional linear regression but bounds the mean response $\mathbb{E}(Y|\x)=\mathbb{P}(Y=1|\x)$, or equivalently the probabilities $p(\x) \equiv \mathbb{P}(Y=1|\x)$, to values between $0$ and $1$. We also assume a different distribution for the response, in this case the binary response or ``labels'' $Y_i$, than in traditional linear regression. Rather than directly fitting $Y_i$ using linear regression with independent and identically distributed (iid) errors $\epsilon_i$ from a normal distribution, we instead assume that the labels $Y_i$ (conditional on $\x_i$) are iid observations from a {\em Bernoulli distribution} with parameter $p_i$. We then fit a linear function to logit transformations of $p_i$; this results in the generalized linear model
\begin{equation}
    \mathrm{logit}(p_i) \equiv \log\bigg(\frac{p_i}{1-p_i}\bigg)={\beta}_0+{\beta}_1x_{i1}+\hdots +{\beta}_dx_{id},\label{eqn:logistic}
\end{equation}
where $x_{ij}$ represents the $j$th predictor or the $j$th component of the input vector $\mathbf{x}_i$. As in standard linear regression, we can use maximum likelihood estimation (MLE) to find the best-fitting ${\bm{\beta}=(\beta_1,\ldots,\beta_d)}$ coefficients. Note that our data come from time series of intensities; due to temporal correlations, the assumption of conditional independence $Y_i|\x_i$ may not hold for events close in time. Nevertheless, our logistic lasso model can still provide useful diagnosis.

The advantage of a linear model, or in this case a generalized linear model, is that it is easy to directly relate predictors to the mean response $\mathbb{E}(Y_i|\x_i)$. Recall that the coefficients in a traditional linear regression model ${Y_i={\beta}_0+{\beta}_1x_{i1}+\hdots +{\beta}_dx_{id}+\varepsilon_i}$ with normally distributed errors $\varepsilon_i$ reflect how important a predictor is relative to the other variables in the model. In a linear model, the regression coefficient $\beta_j$ tells us that an increase in the $j^\mathrm{th}$ variable $x_{ij}$ by one unit, while holding all the other $d-1$ variables constant, is on average associated with an increase (if $\beta_j>0$) or decrease (if $\beta_j<0$) in the response $Y_i$ by $|\beta_j|$ units. The linearity of logistic regression admits a similarly straightforward interpretation that relates the probabilities $p_i$ to the predictors. From Equation \ref{eqn:logistic} we have that an increase of $x_{ij}$ by one unit, holding all other variables constant, is associated with a change of the log-odds (i.e., the logarithm of the odds $\frac{p_i}{1-p_i}$) of the event $Y_i=1$ by ${\beta}_j$, which is equivalent to multiplying the odds by $\exp({\beta}_j)$.

Unfortunately, when $d$ is large (i.e., when we have many predictors), the estimate $\widehat{\bm\beta}$ is highly variable. The statistical model also becomes difficult to interpret. To reduce the variance of the statistical model and to help interpretability, one typically reduces the size of the model by excluding variables that contribute little to the fit from the predictor set. The \emph{logistic lasso} is able to sift through a large number of predictors by maximizing a penalized MLE problem
\begin{equation}
    \widehat{\bm\beta}=\underset{\bm\beta}{\arg\max}\bigg(L(\bm\beta;\mathcal{D})-\lambda\sum_{j=1}^d\lvert\beta_j\rvert\bigg),\label{eqn:lasso}
\end{equation}
where $L(\bm{\beta};\mathcal{D})$ denotes the likelihood of the coefficients ${\bm{\beta}=(\beta_1,\ldots,\beta_d)}$ when observing data ${\mathcal{D}=\{(\x_1,Y_1), \ldots, (\x_N,Y_N)\}}$, the term $\lambda \sum_{j=1}^d\lvert\beta_j\rvert$ is the lasso penalty, and $\lambda \geq 0$ is a tuning parameter. Adding this penalty for $\lambda>0$ discourages large (positive or negative) coefficient values and results in all coefficients shrinking towards 0 as $\lambda$ increases. To penalize all predictors equally, the predictors $\x$ are scaled (to standard deviation 1) and centered (to mean 0). For sufficiently large $\lambda$, small regression coefficients are set to 0. Increasing $\lambda$ leads to smaller statistical models with fewer predictors. We fit our logistic lasso model to TCs between 1998-2009 via ten-fold cross-validation; see Appendix \ref{app:cv} for details. Note that we later in Section~\ref{sec:models}b and c assess the final statistical models (with tuned parameters) on an {\em independent} test set not used in the cross-validation; in our case, all TCs from 2010-2016.

\begin{figure}[b]
    \centering
    \includegraphics[width=\linewidth]{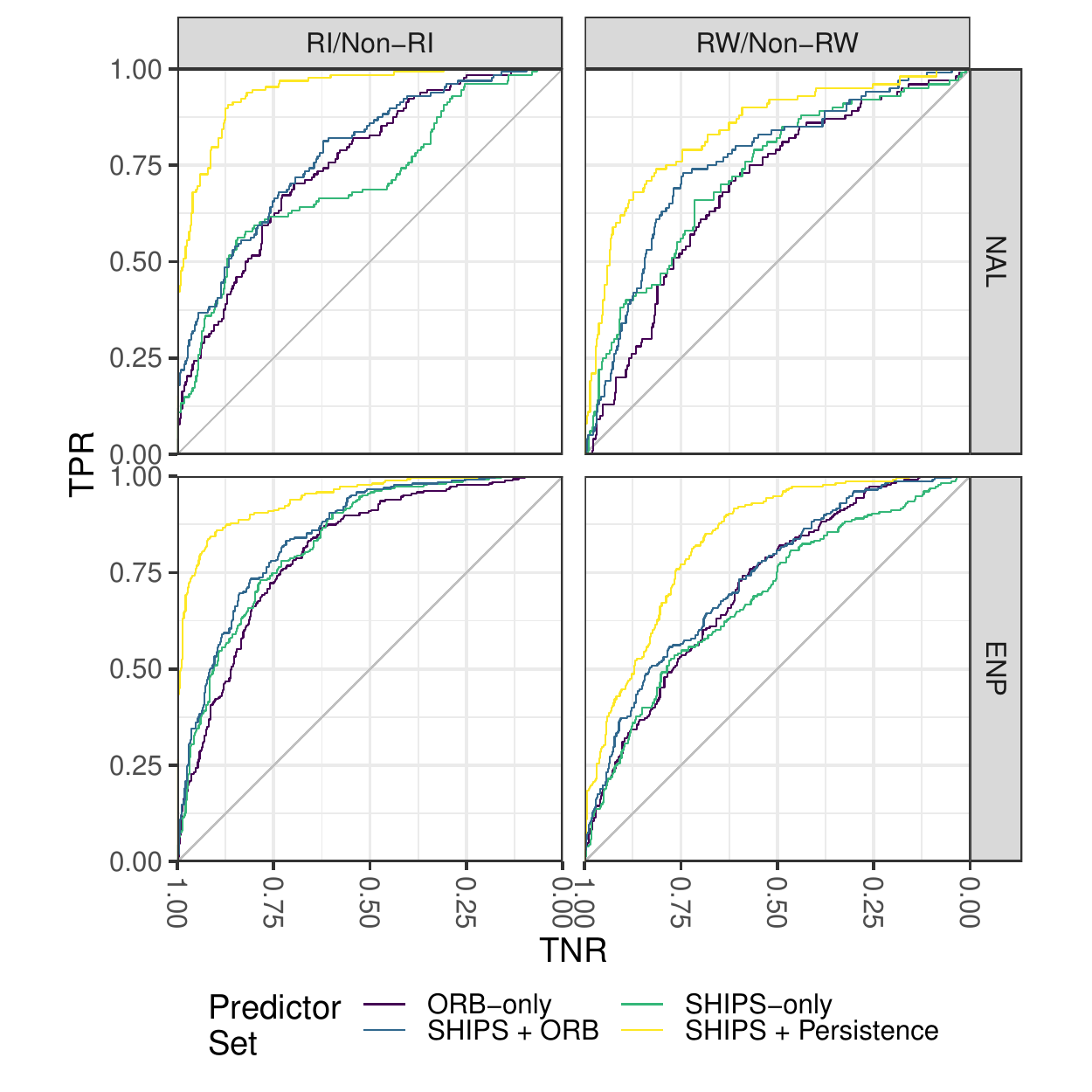}
    \caption{\emph{Classification by logistic lasso}: ROC curve comparison across all 4 predictor sets, by basin and by type of rapid change. See Figure~\ref{fig:AUC_comparison} and Table~\ref{tab:pvals} in Appendix~\ref{app:permute} for AUC metrics with bootstrap confidence intervals and significance tests of differences in AUC between statistical models}.
    \label{fig:roc_comparison}
\end{figure}

{\bf Assessing classification performance.} To evaluate the performance of the fitted statistical models in an objective way, we convert the predicted probabilities $\widehat{p}_i$ to binary classes $C_i$, for which we have a ground truth. Here, $C_i=1$ denotes an ongoing RI (or RW) event and $C_i=0$ denotes the absence of such an event. We perform this conversion by choosing a probability cutoff $p^\star$; for observations $\x_i$ for which the probability $\widehat{p}_i>p^\star$, the classifier predicts onset of or ongoing rapid change ($C_i=1$); if $\widehat{p}_i < p^\star$, the classifier predicts that there is no rapid change ($C_i=0$). {The cut-off} is often fixed to a default value of $p^\star=0.5$, which makes sense in the case of balanced data (i.e., when the two classes occur with the same proportions) but not when one class represents \emph{rare} events such as RI and RW. As we change the value of $p^\star$ in the range $[0,1]$, we can make the resulting classifier more or less sensitive to the event, trading between higher true positive rates at lower $p^\star$ (TPR; the fraction of events correctly identified as such) and higher true negative rates at higher $p^\star$ (TNR; the fraction of non-events correctly identified as such). A so-called receiver operating characteristic (ROC) curve describes the properties of a classifier by showing its TPR and TNR values at different cut-offs $p^\star$ (Figure \ref{fig:roc_comparison}). An ideal classifier would have a ROC curve hugging the top-left corner, which represents a 100\% true positive and true negative rate. (The trivial SHIPS~+~Persistence model tends to perform best, as expected.) A poor classifier would be near the gray diagonal, which corresponds to no better than chance. Hence, a common way of quantifying the performance of a classifier for all possible choices of thresholds is to compute the area under the ROC curve (AUC). In our study, we include both ROC curves and AUC values. In addition, to address the issue of unbalanced data, we choose the threshold $p^\star$ of the binary classifiers in Section~\ref{sec:models}\ref{sec:results} to maximize the so-called ``balanced accuracy'' (equivalent to half of the Peirce score) defined as $BA=(TPR+TNR)/2$ \citep{brodersen2010balanced}.

\begin{figure}[b]
    \centering
    \includegraphics[width=\linewidth]{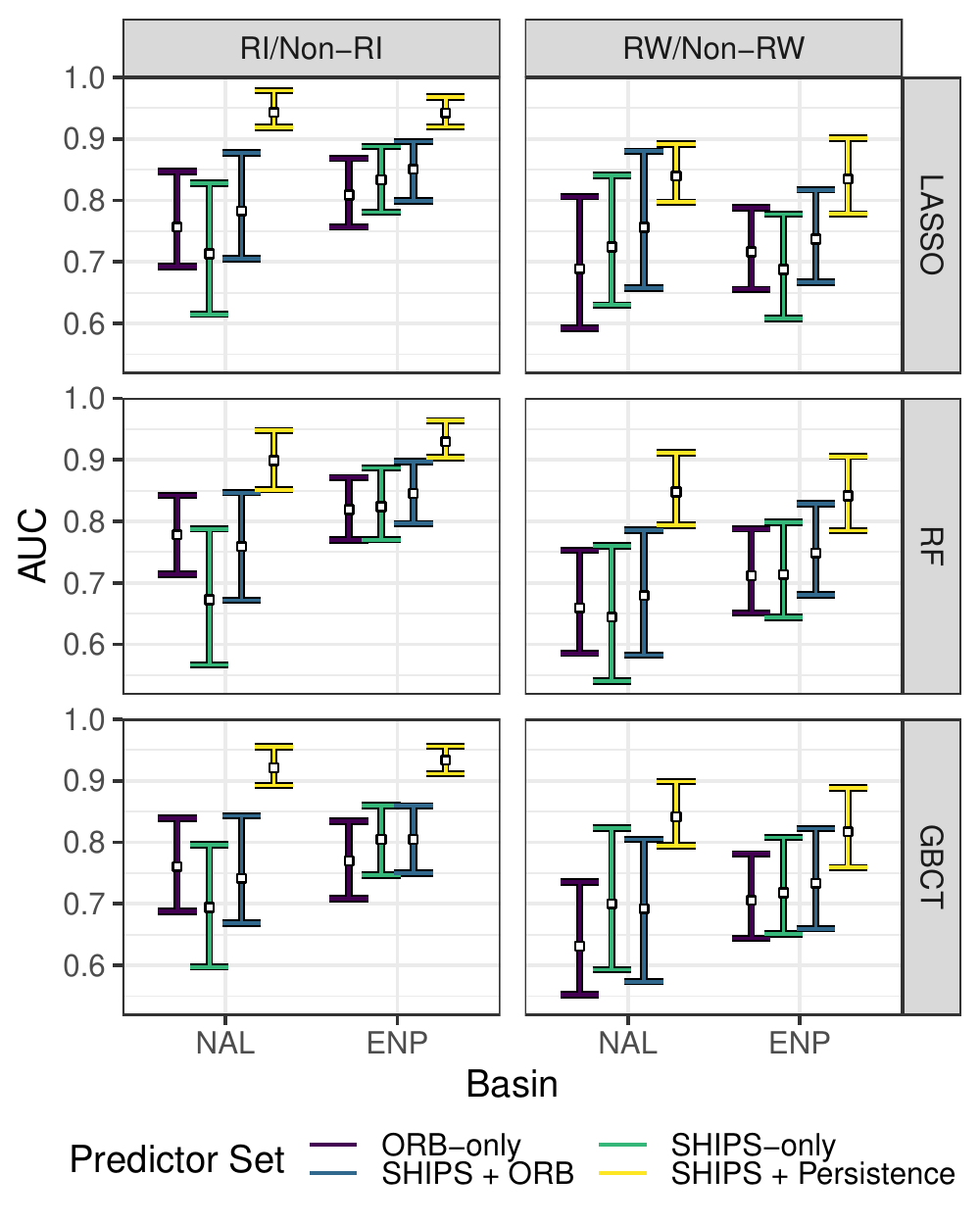}
    \caption{\emph{Binary classification by logistic lasso and nonlinear classifiers:} This plot shows the area under the ROC curve (AUC; white markers) for test data, with bootstrap-estimated 95\% confidence intervals (colored intervals) computed for each of the 16 event--basin--predictor type combinations using the  classifiers ``LASSO'', ``RF'', and ``GBCT''. For all settings, ORB-only has a performance on par with SHIPS-only. Qualitative results persist across different classifiers. See text for details.}
    \label{fig:AUC_comparison}
\end{figure}

\subsection{Results: Logistic Lasso}\label{sec:results}

{\bf Classification by logistic lasso.} We fit the lasso coefficients $\widehat{\beta}$ in Equation \ref{eqn:logistic} for all four predictor sets in Table \ref{tab:predictors}. Figure \ref{fig:roc_comparison} and Figure~\ref{fig:AUC_comparison} (top) summarize the final lasso classification results on test data by type of rapid change (RI or RW) and by basin (ENP or NAL); with four settings and four predictor sets there are 16 different models. In Figure~\ref{fig:AUC_comparison}, the square markers represent the AUC on TCs between 2010-2016 (the test sample) for statistical models fitted on TCs between 1998-2009 (the train sample). We assess the uncertainty in the AUC estimates (conditional on the train sample) by resampling the test sample 250 times with replacement. The figure shows a 95\% pivotal bootstrap confidence interval of the AUC for each model fitted with logistic regression or one of the nonlinear classifiers discussed in Section~\ref{sec:models}\ref{sec:nonlinear}.

\begin{figure*}
    \centering
    \includegraphics[width=.9\linewidth]{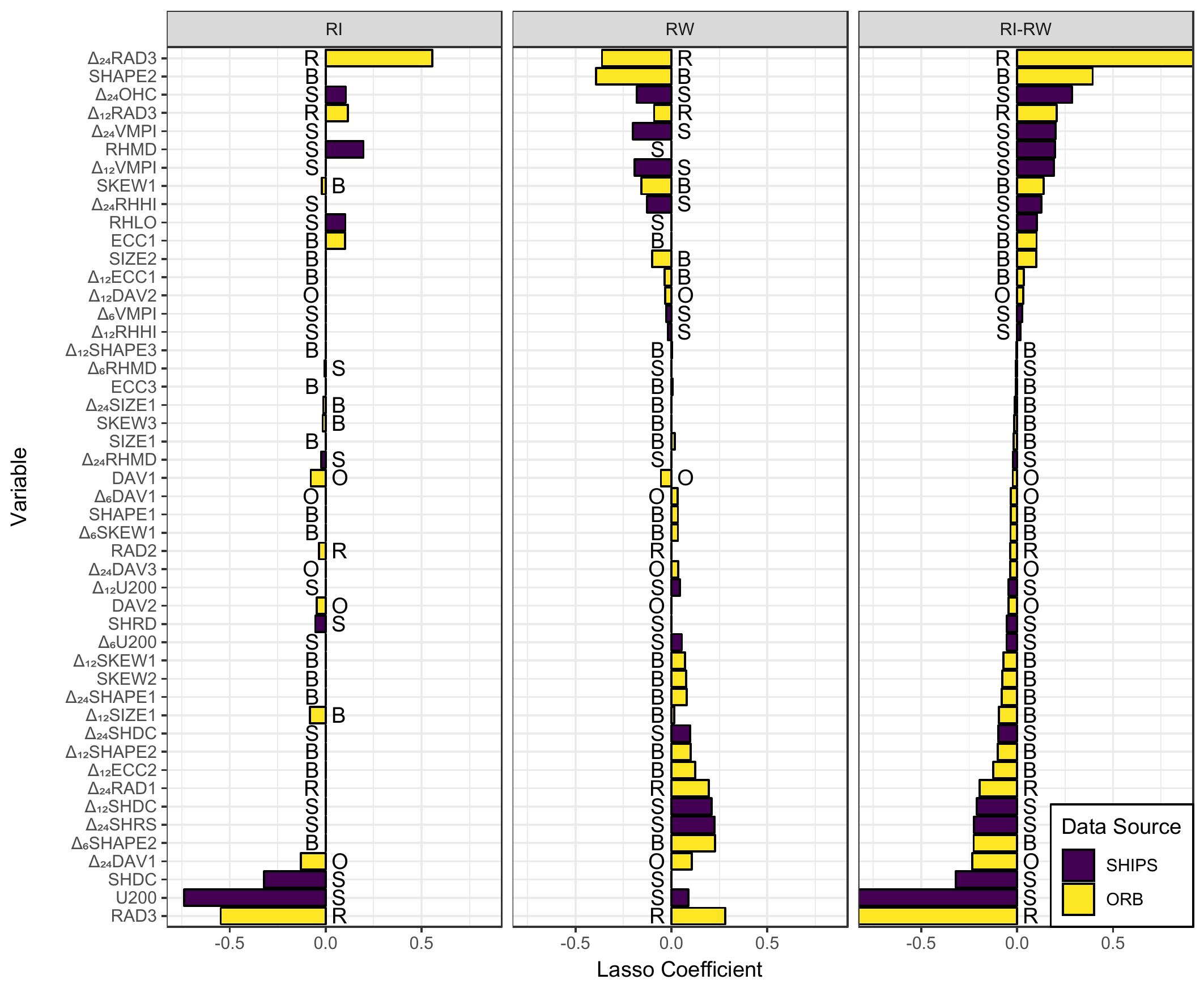}
    \caption{\emph{Interpreting regression coefficients in logistic lasso with the SHIPS~+~\texttt{ORB} predictor set; NAL:} Regression coefficients for SHIPS~+~\texttt{ORB} lasso models for RI events (left) and RW events (center), as well as the difference between the RI and RW regression coefficients in the NAL basin (right). Variables are ordered vertically by RI-RW such that an increase in the variable with the largest RI-RW coefficient, holding all other variables constant, is associated with an increase in the probability of RI much more strongly than an increase in the probability of RW. See text for discussion.}
    \label{fig:lasso_coef_NAL}
\end{figure*}

According to a permutation test\footnote{See Appendix~\ref{app:permute} for details on the permutation testing.} of differences in AUC, the statistical models based on the satellite-derived \texttt{ORB}-only predictor set achieve a classification accuracy comparable to the models based on the SHIPS-only environmental predictor set. This result suggests that \texttt{ORB} is able to adequately assess aspects of evolving convective structure relevant to intensity change by directly capturing the structure of clouds (rather than the environment in which those structures evolve). Furthermore, \texttt{ORB} predictors used as a complement to traditional environmental predictors, as in the SHIPS~+~\texttt{ORB} predictor set, may further improve accuracy. The significance tests for RI/non-RI in the NAL basin in particular underscore the potential for the \texttt{ORB} framework to complement environmental predictors.

In summary, our analysis implies that comprehensive study of GOES IR can provide a high-resolution view of the evolution of TC internal structure, and that information complements traditional NWP-derived environmental predictors. These results appear robust across widely different classification methods (see Section \ref{sec:models}\ref{sec:nonlinear} and Figure \ref{fig:AUC_comparison}).

{\bf Regression coefficients in the fitted logistic lasso.} As mentioned, the logistic lasso model (Equation~\ref{eqn:logistic})  allows for fast variable selection and straightforward interpretation of the regression  coefficients $\widehat{\beta}$. Figure~\ref{fig:lasso_coef_NAL} shows the fitted non-zero lasso coefficients (20 for RI/non-RI and 37 for RW/non-RW) in the NAL basin; the corresponding ``selected'' variables form a subset of the 116 SHIPS~+~\texttt{ORB} predictors. 

The regression coefficient for RAD3 in this model is $-0.53$, meaning that a $1\sigma$ increase in RAD3 (the clarity of the eye-eyewall structure) is associated with a 0.53 decrease in the predicted log-odds of the rapid change event, or a decrease of a factor of $\exp{(-0.53)}=0.59$ in the predicted odds, given that all other predictors are held constant. Likewise, the coefficient for U200 in that same fit is $-0.75$; the same $1\sigma$ increase in U200 is associated with a decrease in the log-odds by 0.75, and the odds by a factor of $\exp{(-0.75)}=0.47$. We center the predictors (in addition to scaling to $\sigma=1$) so that a predictor value of zero corresponds to the sample mean value in that basin.

When predictors with high collinearity are present, a traditional multiple regression will suffer from inflated coefficient variance. Lasso suffers from a different version of the problem, since it will often select one variable out of a group with strongly correlated variables, returning regression coefficients of 0 for the rest. For this reason, the recovered support (the set of predictors with nonzero regression coefficients in the fit) of the final fitted classifier  may exclude predictors expected to appear; e.g., OHC does not appear in the logistic lasso for ENP RI, but that does not mean it is not important to the intensity change process. Instead, other variables (such as RSST) may contain similar information with respect to intensity change as OHC. This effect is reversed for ENP RW, where RSST is dropped from the model while OHC remains.

A positive regression coefficient means that above-average values of a predictor are associated with increased log-odds of rapid intensity change, while negative coefficients mean that above-average values of a predictor are associated with reduced log-odds. Hence, rapid intensity change is more likely when the sign of the predictor matches the sign of the coefficient. In the left panel, the classifier for RI tends to predict ongoing RI events when i) current eye-eyewall structure is weak (negative RAD3 regression coefficient), ii) eye-eyewall structure is strengthening (positive $\Delta_{24}$RAD3 regression coefficient), iii) current 200-hPa zonal winds are low (negative U200 regression coefficient), iv) current wind shear is low (negative SHDC regression coefficient), and v) current mid-level humidity is elevated (positive RHMD regression coefficient). In the center panel, the RW classifier uses a wider range of weaker effects, predicting RW for cases of i) higher current interior symmetry than exterior symmetry (negative SHAPE2 regression coefficient), ii) recent decay of interior symmetry (positive $\Delta_{6}$SHAPE2 and $\Delta_{12}$SHAPE2 regression coefficients), iii) decaying eye-eyewall structure (negative $\Delta_{24}$RAD3 regression coefficient), and iv) strong current eye-eyewall structure (high RAD3 regression coefficient).

The SHIPS~+~\texttt{ORB} results suggest a more complex relationship between structure, environment, and intensity in the case of RW than in the case of RI (via a large number of small to medium-sized regression coefficients as opposed to a few dominant effects). RI is thought to be strongly driven by internal processes, while RW results from a more complex relationship between the TC and its environment. This is reflected in the logistic lasso fitted under the \texttt{ORB}-only setting (shown in supplemental material). RI is dominated by \texttt{R}adial profile predictors (for core structure), SHAPE2 (which reflects the core symmetry), and SKEW1 (for overall skew of the cloud tops, which is influenced by upper-level winds and shear). RW encompasses a wider range of regression coefficients, including global measures such as SIZE, various DAV predictors, and change in overall $T_b$ ($\Delta_{24}$RAD1), alongside the core structure measurements.

The logistic lasso models fitted for the ENP basin (see supplemental material) show similar results, with differences largely explained by the narrow region of TC activity. Outside of this region, SSTs decrease and become hostile to TCs. The logistic lasso for RI fitted under the SHIPS~+~\texttt{ORB} predictor set predicts RI when i) SST is high (RSST), ii) shear is low (SHDC), iii) the TC is farther south (LAT), iv) $T_b$ is low and dropping, particularly in the core (RAD2; RAD1; $\Delta_{24}$RAD1), v) the eye-eyewall structure is intensifying ($\Delta_{24}$RAD3), and vi) the storm has access to ample upper-level moisture (RHHI). 

\subsection{Results: Nonlinear Classifiers}\label{sec:nonlinear}

To determine the robustness of our results across classification methods as well as to benchmark the logistic lasso, we implemented two additional, nonlinear classification methods: random forests (RF) and gradient-boosted classification trees (GBCT). Both RF and GBCT can capture complex relationships between variables, such as interactions between predictors and nonlinear relationships between predictors and rapid change. First, the \texttt{ranger} implementation of random forests (\citealt{breiman2001random,wright2015ranger}) tallies class ``votes'' from many deeply grown classification trees with variables chosen from a random subset at each branch, sampling the training data with replacement for each tree.\footnote{The RF models use Gini impurity for node splitting, with $\sqrt{d}$ predictors sampled at each node. Each forest contains 10,000 trees.} Second, the \texttt{Xgboost} implementation of gradient-boosted classification trees (\citealt{friedman2002stochastic,chen2016xgboost}) iteratively builds trees of a fixed size, fitting the error remaining after the addition of the previous tree.\footnote{The GBCT models use logistic loss and a learning rate of 0.001 on depth-2 trees. The number of trees varies for each model.}

More complex machine learning methods (such as RF, GBCT, and neural networks) have a higher capacity to fit data well but require larger training sample sizes to achieve a better fit compared to simpler linear models (such as the logistic lasso). Indeed, in our settings the logistic lasso either matches or outperforms RF and GBCT (Figure \ref{fig:AUC_comparison} and Table~1 in Supplemental Material). The performance of the two higher-capacity classifiers here appears to be limited by the small size of the historical sample and the few occurrences of rapid changes within that sample. 

With regards to robustness of results across methods: The relative performance of the 16 statistical models fitted for different rapid change types, basins, and predictor sets is consistent across all three classification methods (shown as rows in Figure \ref{fig:AUC_comparison}). In addition, Figures~S7 and S8 of Supplemental material demonstrate that the variable selection and ranking of predictor importances to remain stable across these rather different machine learning methods; for example, the Gini importance in RF tends to agree with the magnitude of the lasso coefficients. Note that RF and GBCT to some extent account for correlation between predictors with different criteria for variable selection. Our analysis indicates that our main results are model-agnostic. We conclude that logistic lasso is a good statistical model for rapid change in our setting because of its statistical performance for the data at hand, fast variable selection, robustness of qualitative results, and added interpretability.

\section{Conclusions and Future Work}

{\bf Conclusions.} High spatial and temporal resolution images from GOES platforms hold promise to support improvements in understanding and forecasting TC intensity change in combination with statistical tools that can ``unlock'' the rich information they contain. The \texttt{ORB} framework provides a method to generate a set of \texttt{ORB} coefficients which target aspects of TC spatial structure in IR imagery that are meaningful to scientists and forecasters and enable both application and interpretation of powerful statistical methods. In this work, we use the logistic lasso --- a generalized linear regression model --- to capture the relationship between IR imagery and rapid change events. The logistic lasso can accept a large number of explanatory variables (including the \texttt{ORB} coefficients from PCA) as inputs and then automatically select a subset of variables relevant to rapid change with fitted regression coefficients that give a measure of the relative importance of the selected inputs.

We apply our \texttt{ORB} framework to GridSat-GOES IR imagery and base the computations on novel (\texttt{B}ulk morphology) as well as existing (\texttt{R}adial profiles and \texttt{O}rganization) \texttt{ORB} statistics. Our approach recovers a low-dimensional phase space for each structural feature, enabling visualization of the trajectory of evolving TC convective structure over time as well as the application of statistical analysis methods such as the logistic lasso.

The \texttt{ORB} coefficients show promise in classifying rapid intensity change events on their own (as in the \texttt{ORB}-only predictor set); however, for best results these features should be used alongside SHIPS predictors. SHIPS and \texttt{ORB} predictors have complementary strengths: SHIPS values provide current estimates of the TC environment, and \texttt{ORB} features capture the history and current structure of the TC itself, which likely contain important markers of ongoing and future changes in TC intensity. In addition, the extended historical record available for both sets of predictors will support future studies and guidance tools in analyzing and leveraging the relationship between the environment and TC structure as revealed by \texttt{ORB}.

Though nonlinear machine learning techniques are capable of fitting more complex relationships, they do not seem to return better predictions for these data. There is a trade-off between higher-capacity, more flexible methods and the need for larger training samples. The number of unique RI and RW events in our study is of the same order as the number of predictors (Table \ref{table:data}). Linear methods place more structural assumptions on the statistical model and tend to perform better in a low-sample-size setting. This property, in combination with variable selection and straightforward interpretation of regression coefficients, makes the logistic lasso an ideal tool to study the relationship between onset of or ongoing rapid change events and environmental and structural predictors.

{\bf Future Work.} The \texttt{ORB} framework is algorithmic, making it easily extensible to other aspects of convective structure in IR imagery and to other bands such as water vapor (WV) or derived products such as differenced IR-WV imagery (\citealt{olander2009tropical}). Such extensions may better probe smaller-scale, transient structures in high-resolution IR imagery and identify spatial and/or temporal patterns in overshooting tops revealed by differenced IR-WV imagery. We expect this framework to be able to extract similar features from the higher-resolution ABI observations, and we will investigate \texttt{ORB}'s performance once more data become available.

Our logistic lasso approach has demonstrated the value of GOES-derived \texttt{ORB} coefficients in a simplified setting: classifying binary RI versus non-RI (or RW versus non-RW) events under the assumption of independence of nearby observations in time. Future work will relax these assumptions by i) treating intensity change as a continuous quantity rather than discretizing it into RI/non-RI or RW/non-RW  and ii) accounting for temporal dependence between observations by creating a time series model which accounts for persistence in environmental and GOES-derived predictors alike. Such statistical models could augment existing intensity guidance schemes by identifying spatial-temporal markers of upcoming intensity change in IR imagery, and provide the means of adding GOES-derived evolutionary history to NWP-derived predictions of a TC's future environment.

Finally, this work utilizes a linear model without interaction terms. Rapid intensity change events are driven by the interplay of internal TC processes and the TC environment; future analyses will examine the effects of such interactions, particularly between \texttt{ORB} coefficients and environmental predictors, by adding interaction terms to the statistical model. As the suite of predictors grows larger and more complex, the logistic lasso can be modified (such as via the group lasso penalty of \citealt{yuan2006grouplasso}) to consider groups of predictors rather than individual predictors, resulting in more intuitive models.

\emph{Acknowledgments.} ABL and TM were supported in part by the National Science Foundation under DMS-1520786, and KW was supported by the Mississippi State University Office of Research and Economic Development. We also thank Dr. David John Gagne for his helpful comments on the manuscript.

\bibliographystyle{ametsoc2014}
\bibliography{references}

\appendix\section{Details on the Calculation of Empirical Orthogonal Functions via Principal Component Analysis}\label{app:pca}

We first center our data so that the sample mean of each \texttt{ORB} function is zero: $\frac{1}{n} \sum_{i=1}^{n} {\bf z}_i = 0$, where a densely sampled \texttt{ORB} function evaluated at $d$ threshold values is here represented by the (random) vector ${\bf z}$, and $n$   denotes the number of observations of ${\bf z}$ in the basin of interest. The centered observations {${\bf z}_1, {\bf z}_2, \ldots {\bf z}_n$} then form the rows of an $n \times d$ matrix ${\bf Z}$. Note that the sample covariance matrix of $\bf z$ is given by the $d \times d$ matrix ${\bf S}= \frac{1}{n}{\bf Z}^T{\bf Z}$.

\begin{figure}
    \centering
    \includegraphics[width=\linewidth]{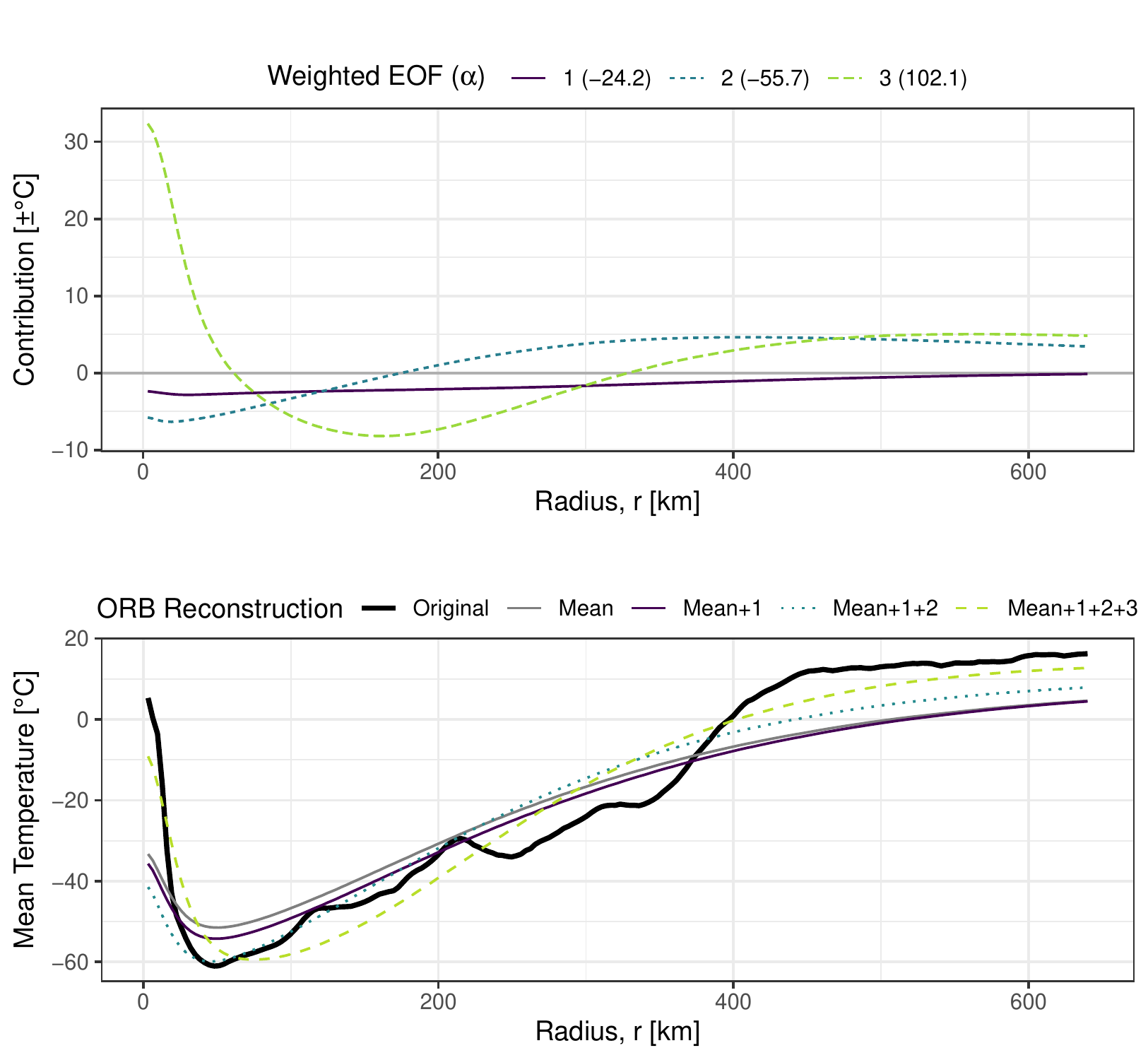}
    \caption{\emph{Reconstructing an \texttt{ORB} function}: The radial profile for Edouard (Figure \ref{fig:radialprofile}) is projected onto the the first three EOFs for radial structure (Figure \ref{fig:radialbasis}), resulting in $K=3$ \texttt{ORB} coefficients $(-24.2, -55.7, 102.1)$. We show the contribution of each EOF to the reconstruction of $\overline{T}(r)$; i.e., the weighted EOFs $\alpha_i f_i(r)$ for $i=1,2,3$ (top). To approximate the original radial profile (bottom; ``Original'', solid black) we add these three contributions to the basin mean (``Mean'', solid gray). With all $K=3$ contributions (``Mean+1+2+3'', green dashed), we find a close approximation to the original function, capturing the overall structure of the observed \texttt{ORB} function $\overline{T}(r)$.}
    \label{fig:rad_recon}
\end{figure}

In principal component analysis (PCA), one computes the eigenvectors of the covariance matrix $\bf Z$. These vectors, which we refer to as empirical orthogonal functions (EOFs), form a basis for the original data. Let ${\bf v}_1, \ldots, {\bf v}_K$ in $\mathbb{R}^d$ denote the eigenvectors that correspond to the $K < d$ largest eigenvalues. In this work, we refer to $v_i$ as the $i^{th}$ EOF, $f_i(x)$. In a principal component (PC) map, the projections of the data onto these vectors are used as new coordinates; i.e., the PC map of the observation ${\bf z}_i$ is given by
\begin{equation}
     {\bf z}_i \mapsto ({\bf z}_i \cdot {\bf v}_1, \ldots, {\bf z}_i \cdot {\bf v}_K)=(\alpha_1,\ldots,\alpha_K),\notag
\end{equation} where the dot represents a scalar product of vectors. In the \texttt{ORB} framework, the scalar projections $(\alpha_1,\ldots,\alpha_K)$ are our \emph{\texttt{ORB} coefficients}. We can reconstruct the approximate \texttt{ORB} function with  \texttt{ORB} coefficients and EOFs according to 
\begin{equation}
    f(x)\approx\overline f(x)+\alpha_1f_1(x)+\ldots+\alpha_Kf_K(x),\notag
\end{equation}
as demonstrated in Figure \ref{fig:rad_recon}. The PC map with \texttt{ORB} coefficients  $(\alpha_1,\ldots,\alpha_K)$ acts as a low-dimensional ``phase space'' for an aspect of convective structure; see Figure~\ref{fig:sizetrajectory} for an example.

Algorithmically, the PC map is easy to compute using a singular value decomposition (SVD) of ${\bf Z}$:
\begin{equation}
 {\bf Z=U D V}^T.\notag
\end{equation}
Here ${\bf U}$ is an $n \times d$ orthogonal matrix,  ${\bf V}$ is a $d \times d$ orthogonal matrix (where the columns are eigenvectors ${\bf v}_1, \ldots, {\bf v}_d$ of ${\bf S}$), and ${\bf D}$ is a $d \times d$ diagonal matrix with diagonal elements $a_1 \geq a_2 \ldots \geq a_d \geq 0$ known as the singular values of ${\bf Z}$. The  first $K$ EOFs are given by the first $K$ columns of ${\bf V}$.  Because ${\bf ZV}={\bf UD}$, the PC embedding of the $i^{th}$ data point in $K$ dimensions 
--- that is, the set of \texttt{ORB} coefficients  $(\alpha_1,\ldots,\alpha_K)$ ---
is given by the first $K$ elements of the $i^{th}$ row of ${\bf UD}$.

\section{Fitting the Logistic Lasso}\label{app:cv}

We fit the logistic lasso model on TCs prior to 2010 using the \texttt{glmnet} package for \texttt{R} \citep{friedman2009glmnet}; we select $\lambda$ via tenfold cross-validation (CV). This procedure fits ${\bm\beta}$ on nine out of ten so-called ``folds,'' holding out a rotating tenth so-called validation fold. The tuning parameter $\lambda$ is chosen so as to maximize the fit on the the held-out validation folds (see \citealt{hastie2005cv} for details).  CV prevents ``overfitting'' (fitting too closely to training data, hence performing poorly on future observations) or ``underfitting'' (inadequately capturing generalizable features in the training data). In our setting, CV results in a different $\lambda$-value for each of the 16 fits. When choosing CV folds, we randomly sample TCs rather than observations in order to account for temporal dependence within a single TC's lifespan. Note that we {verify the final statistical models} (with tuned parameters) on an independent test set {\em not} used in the cross-validation; in our case, this test set consists of all TCs from 2010-2016.

\section{Testing for Differences in AUC for Fitted Statistical Models}\label{app:permute}

{\bf Hypotheses.} We formally test two of the conjectures stated in the introduction to this work.
\paragraph{Test 1:} Our first conjecture is that \texttt{ORB}-only does at least as well as SHIPS-only. In words, we ask ``Is there evidence that \texttt{ORB}-only would do worse than SHIPS-only?''. That is, test 
\begin{equation}
\begin{split}
H_0:\ AUC_{\text{\texttt{ORB}-only}}&=AUC_{\text{SHIPS-only}}\\  \text{versus}\ H_a:\ AUC_{\text{\texttt{ORB}-only}}&<AUC_{\text{SHIPS-only}}.
\end{split}
\end{equation}

\paragraph{Test 2:} Our second conjecture is that SHIPS~+~\texttt{ORB} may improve upon SHIPS-only. In words, we ask  ``Is there evidence that SHIPS~+~\texttt{ORB} improves upon SHIPS-only?''. That is, test
\begin{equation}
\begin{split}
    H_0:\ AUC_{\text{SHIPS~+~\texttt{ORB}}}&=AUC_{\text{SHIPS-only}}\\ \text{versus}\ H_a:\ AUC_{\text{SHIPS~+~\texttt{ORB}}}&>AUC_{\text{SHIPS-only}}.
\end{split}
\end{equation}
 
\begin{table}[b]
\ra{1.2}
\centering
\begin{tabular}{lccccccc}
& \multicolumn{3}{c}{RI/non-RI} && \multicolumn{3}{c}{RW/non-RW} \\
\cmidrule{2-4} \cmidrule{6-8}
      &  NAL  &&  ENP  &&  NAL  &&  ENP\\
\cmidrule{2-2} \cmidrule{4-4} \cmidrule{6-6} \cmidrule{8-8}
LASSO & 0.886 && 0.151 && 0.232 && 0.828\\
RF    & 0.999 && 0.403 && 0.639 && 0.449\\
GBCT  & 0.952 && 0.079 && 0.065 && 0.323\\
& \multicolumn{7}{r}{$H_0:\ AUC_{\text{\texttt{ORB}-only}}=AUC_{\text{SHIPS-only}}$}\\
& \multicolumn{7}{r}{$H_a:\ AUC_{\text{\texttt{ORB}-only}}<AUC_{\text{SHIPS-only}}$}\\
\hline
LASSO & \bf{0.026} && 0.175 && 0.191 && \bf{0.048}\\
RF    & \bf{0.009} && 0.156 && 0.230 &&     0.099\\
GBCT  &     0.105  && 0.491 && 0.596 &&     0.291\\
& \multicolumn{7}{r}{$H_0:\ AUC_{\text{SHIPS~+~\texttt{ORB}}}=AUC_{\text{SHIPS-only}}$}\\
& \multicolumn{7}{r}{$H_a:\ AUC_{\text{SHIPS~+~\texttt{ORB}}}>AUC_{\text{SHIPS-only}}$}\\
\hline
\end{tabular}
\caption{P-values for testing for differences in AUC relevant to the goals of the paper, estimated as in Appendix~\ref{app:permute}. P-values significant at the 0.05 level for each individual model are shown in bold. Rows 1-3 correspond to tests of \texttt{ORB}-only AUCs lagging significantly behind SHIPS-only, while rows 4-6 correspond to tests of SHIPS~+~\texttt{ORB} improving on SHIPS-only.}
\label{tab:pvals}
\end{table}
 
{\bf Test Statistics.} Let $X_1,...,X_N$ be the  probabilistic predictions from the \texttt{ORB}-only model for test data, and let $Y_1,...,Y_N$ be the predictions from the SHIPS-only model for the same $N$ test points. Our test statistic is the difference in AUC values between the two statistical models,
\begin{equation}
\begin{split}
    T_1(X_1,...,X_N,Y_1,...,Y_N)&=\\AUC(X_1,&...,X_N)-AUC(Y_1,...,Y_n).
\end{split}
\end{equation}

Likewise for the second test, let $X_1,...,X_N$ be the probabilistic predictions from the SHIPS~+~\texttt{ORB} model for the same $N$ test data, and $Y_1,...,Y_N$ be the predictions for SHIPS-only as before. Then, define the second test statistic as
\begin{equation}
\begin{split}
    T_2(X_1,...,X_N,Y_1,...,Y_N)&=\\AUC(X_1,&...,X_N)-AUC(Y_1,...,Y_N).
\end{split}
\end{equation}

{\bf Permutation Tests.} To assess the significance of an observed difference in AUC, we estimate the distribution of $T_i$ under the null by forming $B=1,000$ permutations of the sample points. That is, for $b=1,\ldots,B$, randomly sample  $\Tilde{X}^b_1,...,\Tilde{X}^b_N$ from $\{X_1,...,X_N,Y_1,...,Y_N\}$ without replacement, and assign the remaining $N$ predictions to $\Tilde{Y}^b_1,...,\Tilde{Y}^b_N$. For each of the B permutations, compute $\Tilde{T}^b_i=T_i(\Tilde{X}^b_1,...,\Tilde{X}^b_N,\Tilde{Y}^b_1,...,\Tilde{Y}^b_N)$. Our estimated p-values for the two tests are given by
\begin{equation}
    \hat{p}_1=\frac{1}{B}\sum^B_{b=1}\mathbb{I}(\Tilde{T}_1^b<T_1)\ \text{and}\ \hat{p}_2=\frac{1}{B}\sum^B_{b=1}\mathbb{I}(\Tilde{T}_2^b>T_2),
\end{equation}
where $\mathbb{I}(\cdot)$ is the indicator function, taking value $1$ when the statement is true and value $0$ when it is false.

\newpage
\onecolumn
\section{Additional Figures}

\begin{figure*}[h]
    \centering
    \includegraphics[width=.7\linewidth]{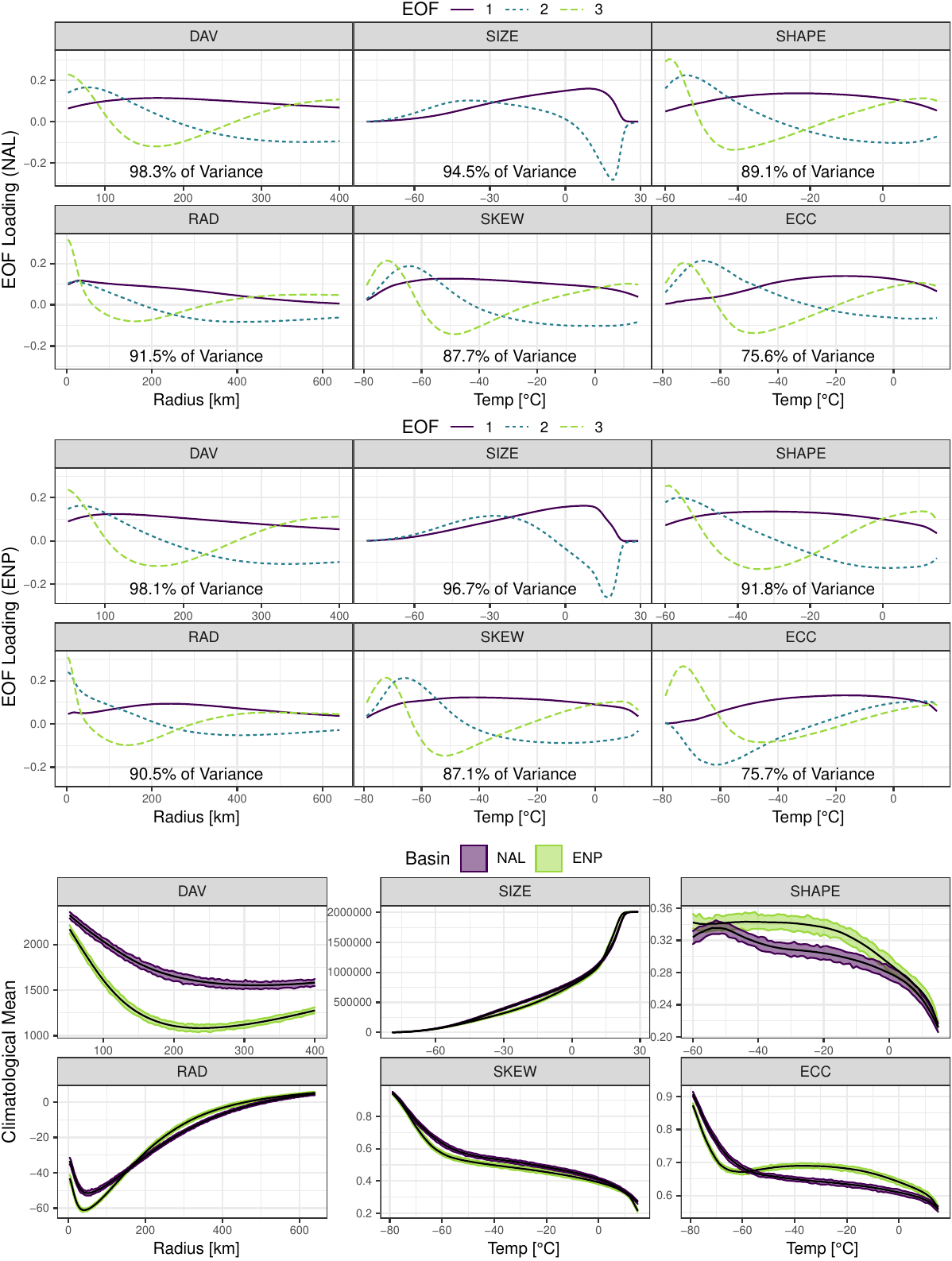}
    \caption{\emph{EOFs and mean ORB Functions:} EOFs for all 6 \texttt{ORB} functions in the NAL (top) and ENP (center), with sample means (bottom) with 95\% point-wise confidence intervals estimated by a stationary bootstrap \citep{politis1994stationary}. The computed EOFs differ significantly only in the sign of ECC2, which merely flips the direction of interpretation. The difference between basins is largely contained in the sample means. In spite of these differences, the same qualitative structures persist between basins.}
    \label{fig:orb_basis}
\end{figure*}

\begin{figure*}[h]
    \centering
     \includegraphics[width=.9\linewidth]{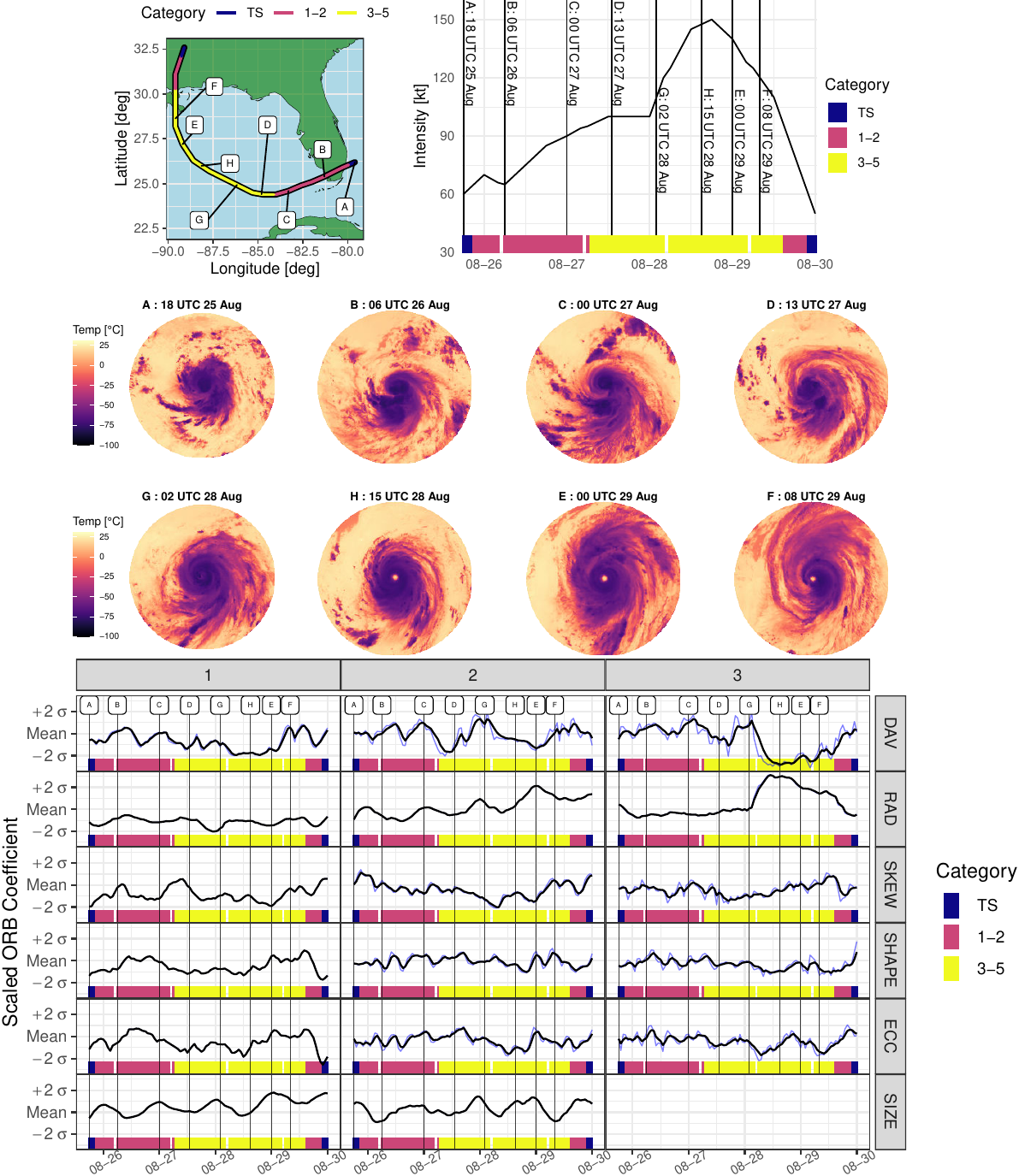}
    \caption{\emph{Full \texttt{ORB} evolution of Katrina (2005) with selected stamps:} A map shows the trajectory of Katrina (top-left) with select points (A-H) labeled. The intensity over time is shown with the same points labeled (top-right). GOES imagery is shown for the 8 labeled points (center). Finally, the full suite of \texttt{ORB} coefficients is shown as a function of time (bottom), with the original time series (\emph{blue}) smoothed by an exponential weighted moving average (EWMA, \emph{black}) with decay chosen to achieve a roughness (average value of $\lvert\sfrac{d^2f(t)}{dt^2}\rvert$) of $0.2\sigma/$hr$^2$ across the basin. Where no blue curve is visible, the smoothed time series nearly equals the original time series.}
    \label{fig:katrina_evolution}
\end{figure*}

\end{document}